\documentclass[english]{eptcs}

\usepackage{times}
\usepackage[latin1]{inputenc}
\usepackage{pslatex}
\usepackage{amsmath}
\usepackage{amssymb}
\usepackage{amsfonts}
\usepackage{amsthm}
\usepackage{hyperref}
\usepackage{breakurl}
\usepackage{prooftree}
\usepackage{txfonts}

\usepackage{color}

\newif\ifmc
\mctrue

\newtheorem{fact}{Fact}[section]

\newtheorem{lemma}[fact]{Lemma}

\newtheorem{corollary}[fact]{Corollary}

\newtheorem{definition}[fact]{Definition}

\newtheorem{theorem}[fact]{Theorem}

\newcommand{\dup}{\! : \!}
\newcommand{\spaziot}{\vspace{-4pt}}
\newcommand{\spazios}{\vspace{-4pt}}
\newcommand{\spaziop}{\vspace{-9.5pt}}
\newcommand{\fhp}{{\large{\textsc{fhp}}}}
\newcommand{\fhi}{{\large{\textsc{fhi}}}}
\newcommand{\tA}{\sigma}       
\newcommand{\tB}{\tau}
\newcommand{\tC}{\rho}
\newcommand{\tD}{\theta}
\newcommand{\tE}{\vartheta}
\newcommand{\tF}{\zeta}
\newcommand{\tG}{\varsigma}
\newcommand{\tM}{\mu}
\newcommand{\tN}{\nu}
\newcommand{\tl}{\lambda}
\newcommand{\te}{\xi}
\newcommand{\tv}{\eta}
\newcommand{\ti}{\chi}
\newcommand{\tk}{\kappa}
\newcommand{\tx}{\iota}
\newcommand{\tu}{\omega}
\newcommand{\tr}{\upsilon}
\newcommand{\tO}{\alpha}
\newcommand{\tP}{\beta}
\newcommand{\tQ}{\gamma}
\newcommand{\tY}{\delta}
\newcommand{\tI}{\alpha}
\newcommand{\tL}{\beta}
\newcommand{\tV}{\eth}
\newcommand{\B}{\Gamma}

\newcommand{\db}{\displaystyle}

\newcommand{\tS}{\sigma}       
\newcommand{\tT}{\tau}
\newcommand{\tR}{\rho}

\newcommand{\der}[3]{#1\vdash#2\!\dup\!#3}
\newcommand{\p}{{\sf p}}

\newcommand{\calP}{{\cal P}}
\newcommand{\calL}{{\cal L}}
\newcommand{\calR}{{\cal R}}

\newcommand{\red}{\Longrightarrow}

\newcommand{\vf}{\varphi}
\newcommand{\w}[1]{\ensuremath{dw(#1)}}
\newcommand{\cw}[1]{\ensuremath{cw(#1)}}

\newcommand{\set}[1]{\ensuremath{\{#1\}}}

\newcommand{\ag}[2]{#1\propto#2}
\newcommand{\C}[1]{{\cal C}[#1]}
\newcommand{\pa}[1]{s(#1)}
\newcommand{\pad}[1]{d(#1)}
\newcommand{\Cp}[1]{{\cal C'}[#1]}
\newcommand{\nag}[2]{#1\not\propto#2}

\newcommand{\id}{{\sf Id}}
\newcommand{\tob}{\;{_\beta}\!\!\longleftarrow}

\newcommand{\labelx}[1]{\label{#1}}    

\newcommand{\rv}[2]{#1\upharpoonright FV(#2)}
\newcommand{\lp}{\swarrow}
\newcommand{\rp}{\searrow}
\newcommand{\ep}{\square}
\newcommand{\q}{{\sf p}}
\newcommand{\Sp}{e}
\newcommand{\Ip}{\#}
\newcommand{\isot}{\approx}

\newcommand{\mi}{\leq^\wedge\!\!}
\newcommand{\muu}{\leq^\vee\!\!}

\newcommand{\md}{\leq^{\Diamond}\!\!}

\providecommand{\event}{ITRS 2012} 

\title{Toward Isomorphism of Intersection and Union Types\footnote{
 This work was partially supported by MIUR Project CINA and Ateneo/CSP Project SALT. 
}\\
{\small \it Dedicated to Corrado B\"ohm on the occasion of his 90th Birthday}}
\author{Mario Coppo\quad
Mariangiola Dezani-Ciancaglini\quad
Ines Margaria\quad
Maddalena Zacchi
\institute{Dipartimento di Informatica Universit\`a di Torino,
corso Svizzera 185, 10149 Torino, Italy}
}

\def\titlerunning{Toward Isomorphism of Intersection and Union types
}
\def\authorrunning{M. Coppo,  M. Dezani-Ciancaglini, I. Margaria \& M. Zacchi 
}
\begin{document}

\maketitle

%

\begin{abstract}
 This paper investigates type isomorphism in a $\lambda$-calculus with intersection and union types. It is known that in $\lambda$-calculus, the isomorphism between two types is realised by a pair of  terms inverse one each other. Notably, invertible terms are linear terms of a particular shape, called finite hereditary permutators. Typing properties of finite hereditary permutators are then studied in a relevant  type inference system with intersection and union types for linear terms. In particular, an isomorphism preserving reduction between types is defined.  Reduction of types is confluent and terminating, and induces a notion of
 normal form of types. The properties of normal types are a crucial step toward the complete characterisation of type isomorphism. The main results of this paper are, on one hand, the fact that two types with the same normal form are isomorphic, on the other hand, the characterisation of the isomorphism between types in normal form, modulo isomorphism of arrow types.
\end{abstract}

\section{Introduction}\spaziot
In a calculus with types, two types $\sigma$  and $\tau$
are \textit{isomorphic} if there exist two terms $P$ of
type $\sigma\rightarrow\tau$ and $P'$ of type
$\tau\rightarrow\sigma$ such that both their compositions $P \circ
P' = \lambda x. P(P'x)$ and $P' \circ  P  = \lambda x. P'(Px) $ give the identity (at the proper type). The study of type
isomorphism started in the
1980s
with the aim of finding 
all the type
isomorphisms valid in every model of a given language \cite{BruceLongo85}.
If one looks at this problem choosing as language
a  $\lambda$-calculus with types, one can immediately note the close
relation between  type isomorphism and  $\lambda$-term
invertibility. Actually, in the untyped $\lambda$-calculus a
$\lambda$-term $P $ is \textit{invertible} if there exists a
$\lambda$-term $P'$ such that $P \circ  P' =_{\beta\eta} P'\circ P =_{\beta\eta} \textbf{I}$
($\mathbf{I} = \lambda x.x$). The problem of
term invertibility has been extensively studied for the untyped
$\lambda$-calculus since 1970 and the main result has
been the complete characterisation of the invertible
$\lambda$-terms in $\lambda \beta \eta$-calculus
\cite{Dezani}: the invertible terms are all and only the
\textit{finite hereditary permutators}.

\begin{definition}
 [Finite Hereditary Permutator] A \emph{finite
hereditary permutator} (\fhp\ for short) is a $\lambda$-term of the form (modulo $\beta$-conversion)
\[\lambda x y_{1}\dots y_{n}.x(P_1 y_{\pi(1)})\dots (P_n y_{\pi (n)}) \; \; \;(n \geq  0)\]
where  $\pi$ is a permutation of $1,\dots,n$, and $P_1,\ldots,P_n$ are \fhp s.
\end{definition}
\noindent
Note that the identity  is trivially an \fhp\ (take $n=0$). Another example of an \fhp\ is
 \\
\centerline{$\lambda x y_1 y_2. x\,y_2\,y_1 = \lambda x y_1 y_2. x\,((\lambda z. z)\,y_2)\,((\lambda z. z)\,y_1).$} 
It is easy to show that \fhp s are closed by composition.

\begin{theorem}
 A $\lambda$ -term is invertible iff it
is a finite hereditary permutator.
\end{theorem}

This result,  obtained in the framework of the untyped
$\lambda$-calculus, has been the basis for studying type
isomorphism in different type systems for the $\lambda$-calculus.
Note that
every \fhp\ has, modulo $\beta \eta$-conversion, a unique
inverse $P^{-1}$.
Even if in the type free $\lambda$-calculus \fhp s are defined in \cite{Dezani} modulo $\beta\eta$-conversion, in this paper each \fhp\ is considered only modulo $\beta$-conversion, because types are not invariant under $\eta$-expansion.
Taking into account these properties, the definition of type
isomorphism in a $\lambda$-calculus with types can be
stated as follows:
\begin{definition}
 [Type isomorphism] Given a  $\lambda$-calculus with types,
two types $\sigma$  and  $\tau$ are isomorphic ($
\sigma \approx \tau$) if there exists a pair $<P,P^{-1}>$ of \fhp s, inverse of each other, such that $\vdash
P\dup \sigma \rightarrow \tau$ and $\vdash P^{-1}
\dup \tau\rightarrow\sigma$. The pair
$<P,P^{-1}>$
\textit{proves} the isomorphism.
\end{definition}
\noindent
When $P=P^{-1}$
one can simply write ``$P$ proves the isomorphism".

The main approach used to characterise type isomorphism in a given
 system has been to provide a suitable  set of equations  and
to prove that these equations  induce the type isomorphism w.r.t.
$\beta\eta$-conversion, i.e. that the types of the \fhp s are all and
only those induced by the  set of equations.

The typed $\lambda$-calculus studied first has been the simply
typed $\lambda$-calculus. For this calculus Bruce and Longo proved in \cite{BruceLongo85}
that only one equation is needed, namely the {\bfseries swap}
equation:
\[\sigma \rightarrow \tau \rightarrow \rho  ~\approx~  \tau \rightarrow \sigma\rightarrow
    \rho\]
Later, the study has been directed toward richer
$\lambda$-calculi, obtained from the simply typed
$\lambda$-calculus  in an incremental way, by adding some other
type constructors (like product and unit types \cite{Solovev83,BruceDicosmoLongo92,soloviev93complete}) or by allowing higher-order types (System F \cite{BruceLongo85,Dicosmo93}).
Di Cosmo summarised in
\cite{MSCSSurvey05} the
equations characterising  type isomorphism;  the set of equations
grows incrementally in the sense that the set of equations
for a typed $\lambda$-calculus, obtained by adding a primitive to a
given $\lambda $-calculus, is an extension of the set
of equations of the $\lambda$-calculus without that primitive.

In the presence of intersection, this incremental approach
does not work, as pointed out in \cite{DDGT10}; in particular  with intersection types, the isomorphism
is no longer a  congruence and type equality in the standard
models of intersection types does not entail type isomorphism.
These quite unexpected facts required the introduction of a syntactical notion of type similarity in order to fully characterise the isomorphic types \cite{DDGT10}.

The study of  isomorphism looks even harder for type systems with
intersection and union types because for these systems, in general, the
Subject Reduction property does not hold \cite{barba}.
As in the case of intersection types, the isomorphism of union types is not a congruence and it is not complete for type equality in standard models. For example $\tA\vee\tB \to \tC$ and $\tB\vee\tA \to \tC$ are isomorphic, while $(\tA\vee\tB \to \tC)\vee\varphi$ and $(\tB\vee\tA \to \tC)\vee\varphi$ are not isomorphic, whenever $\varphi$ is an atomic type.\label{page2}

This paper gives   essential results  for  the characterisation of isomorphism of intersection and union types.
To this aim a relevant type system, defined as a slight modification of the standard one in \cite{MQPS86}, has been introduced. In this system, in particular, Subject Conversion holds for linear terms.

A main difficulty in studying types for \fhp s is that intersection/union introduction and elimination rules allow to write types in different, although isomorphic, ways. Since the standard distributive laws of union and intersection correspond to provable isomorphisms, types can be considered  both in disjunctive and in conjunctive normal forms. This, besides providing a very useful technical tool, allows one to define, together with other basic isomorphisms involving the $\to$ type constructor, a general notion of normal form of types. A main result proved in this paper is that if $\tA\to\tB$ is a type of an \fhp\ $P$, then:
\begin{itemize}
\item for all $\tM$ in the disjunctive normal form of $\tA$, there is $\tN$ in the disjunctive normal form of $\tB$ such that $\tM\to\tN$ is a type of $P$;
\item for all $\tk$ in the conjunctive normal form of $\tB$, there is $\ti$ in the conjunctive normal form of $\tA$ such that $\ti\to\tk$ is a type of $P$.
\end{itemize}

Another crucial contribution is the introduction of normalisation rules which allow to split arrows over intersections/unions and to erase ``useless" types  by preserving isomorphism. The proof of soundness of these rules is done by building the pairs of \fhp s witnessing isomorphism. Termination and confluence of the normalisation rules are also shown. Two types with the same normal form are  isomorphic. Normal types are intersections of unions of atomic and arrow types. A key property is that two isomorphic normal types have the same number of intersections and unions and that the arrows and atoms are pairwise isomorphic. The last step toward a complete characterisation of type isomorphism is that of combining  normal forms with the \textbf{swap} equation,
and this is done in \cite{CDMZ13}.

\section{Type assignment system}\spaziot
\begin{figure}[h]
\centerline{$
\begin{array}{ll@{~~~~~~}ll}
\multicolumn{4}{c}
{(Ax)\quad\quad   x\dup\tA \vdash x\dup\tA} \\
\\
(\to I) &  \db \frac{\B,x\dup\tA \vdash M\dup\tB} {\B \vdash \lambda
x.M\dup\tA \to \tB} & (\to E) & \db \frac{\B_{1} \vdash M\dup\tA \to
\tB \quad
\B_{2} \vdash N\dup\tA}{\B_{1}, \B_{2} \vdash MN\dup\tB} \\
\\
(\wedge I) &  \db \frac{\B \vdash M\dup\tA ~~ \B \vdash M\dup\tB}
{\B \vdash M\dup\tA \wedge \tB} &
(\wedge E) & \db \frac{\B \vdash M\dup\tA \wedge \tB}{\B \vdash M\dup\tA}
~~~~~~\db \frac{\B \vdash M\dup\tA \wedge \tB}{\B \vdash M\dup\tB}\\
\\
\multicolumn{4}{c}{(\vee I) \quad  \db \frac{\B \vdash M\dup\tA}
{\B \vdash M\dup\tA\vee\tB}~~~~~~  \db \frac{\B \vdash M\dup\tA}
{\B \vdash M\dup\tB\vee\tA}}\\\\
\multicolumn{4}{c}{(\vee E) \quad \db \frac{\B_{1}, x\dup\tA\wedge\tD
\vdash M\dup\tC\quad \B_{1}, x\dup\tB\wedge\tD \vdash M\dup\tC\quad
\B_{2} \vdash N\dup(\tA\vee\tB)\wedge\tD}{\B_{1},\B_{2} \vdash
M[N/x]\dup\tC}}
\end{array}
$}
\caption{Typing rules.}\labelx{tr}
\end{figure}

\indent The  syntax of intersection and union types is given by:
\begin{center}
$\begin{array}{lll}
\tA&::=&\varphi~~\mid~\sigma\to\sigma~\mid~\sigma\wedge\sigma~\mid~\sigma\vee\sigma
\end{array}$
\end{center}
\noindent
where $\varphi$  denotes an atomic type.
It is useful to distinguish between different kinds of types. So in the following:
\begin{itemize}
\item $\tA , \tB  , \tC, \tD,\tE,\tF, \tG$  range over arbitrary types;
\item $\tO, \tP, \tQ,\tY,\tV$  range over atomic  and arrow types, defined as ~~$\tO~::=~\varphi~\mid~\tA\to\tA$;
\item $\tM,\tN,\tl,\te,\tv$  range over intersections of atomic  and arrow types ({\em basic intersections}), defined as ~~$\tM~::=~\tO~\mid~\tM\wedge\tM$;
\item $\ti,\tk,\tx,\tu$ range over unions of atomic  and arrow types ({\em basic unions}), defined as ~~$\ti~::=~\tO~\mid~\ti\vee\ti$.\label{basic}

\end{itemize}
Note that no structural equivalence is assumed between types, for instance $\tA \vee \tB$ is different from $\tB \vee \tA$.
As usual, parentheses are omitted according to the precedence rule ``$\vee$   and  $\wedge$ over $\rightarrow$" and 
$\to$ associates to the right.

The union/intersection type system considered in this paper is a
modified version of the basic one introduced in the seminal paper
\cite{MQPS86}, restricted to linear $\lambda$-terms. A $\lambda$-term is {\em linear} if each free or bound variable occurs exactly once in it.

\medskip

Figure
\ref{tr} gives the typing rules. As usual, {\em environments} associate
variables to types and  contain at most one type for each variable.
The environments are relevant, i.e. they contain only the used premises. When writing $\Gamma_1,  \Gamma_2$ one convenes that the sets of
variables in $\Gamma_1$ and  $\Gamma_2$ are disjoint.

The only
non-standard rule is $(\vee E)$. This rule  takes into account the fact that, as it seems natural in a system with intersection types, one variable can be used  in a deduction with different types in different occurrences (by applications of the $(\wedge E)$ rule). It should then be possible, in general, to apply the union elimination only to the type of one of these occurrences. A paradigmatic example is the one in Figure \ref{ex2} where one occurrence of the variable $y$ is used (after an application of $(\wedge E)$) with type $\tA$  in one branch of the $(\vee E)$ rule and with type $\tB$ in the other branch. Other occurrences of $y$ are used instead with type $\tC$ in both branches. Rule $(\vee E)$ is then the right way to formulate union elimination in a type system in which union and intersection interact. It is indeed a generalisation of the $(\vee E')$ rule  given in \cite{MQPS86}.
A last observation is that, being $M$ linear, in an application of the $(\vee E)$ rule, exactly one occurrence of $x$ is replaced inside $M$.

\begin{figure}
{\scriptsize
\begin{prooftree}
\begin{prooftree}
\begin{prooftree}
\begin{prooftree}
[y\dup\tA\wedge\tC]
\justifies
y\dup\tC
\using (\wedge E)
\end{prooftree}
\begin{prooftree}
[y\dup\tA\wedge\tC]
\justifies
y\dup\tA
\using (\wedge E)
\end{prooftree}
\justifies
y\dup\tC\wedge\tA
\using (\wedge I)
\end{prooftree}
\justifies
y\dup(\tC\wedge \tA)\vee (\tC \wedge \tB)
\using (\vee I)
\end{prooftree}
\begin{prooftree}
\begin{prooftree}
\begin{prooftree}
[y\dup\tB\wedge\tC]
\justifies
y\dup\tC
\using (\wedge E)
\end{prooftree}
\begin{prooftree}
[y\dup\tB\wedge\tC]
\justifies
y\dup\tB
\using (\wedge E)
\end{prooftree}
\justifies
y\dup\tC\wedge\tB
\using (\wedge I)
\end{prooftree}
\justifies
y\dup(\tC\wedge \tA)\vee (\tC \wedge \tB)
\using (\vee I)
\end{prooftree}
\begin{prooftree}
\begin{prooftree}
[x\dup\tC\wedge (\tA\vee \tB)]
\justifies
x\dup\tA\vee \tB
\using (\wedge E)
\end{prooftree}
\begin{prooftree}
[x\dup\tC\wedge (\tA\vee \tB)]
\justifies
x\dup\tC
\using (\wedge E)
\end{prooftree}
\justifies
x\dup(\tA\vee \tB)\wedge \tC
\using (\wedge I)
\end{prooftree}
\justifies
\begin{prooftree}
x\dup(\tC\wedge \tA)\vee (\tC \wedge \tB)
\justifies
\lambda x.x\dup \tC\wedge (\tA\vee \tB) \to (\tC\wedge \tA)\vee (\tC \wedge \tB)
\using (\to I)
\end{prooftree}
\using (\vee E)
\end{prooftree}
}
\caption{A deduction of $\vdash \lambda x.x\dup \tC\wedge (\tA\vee \tB) \to (\tC\wedge \tA)\vee (\tC \wedge \tB)$.}\labelx{ex2}
\end{figure}

Some useful admissible rules are:
\[( L) \quad \db \frac{x\dup\tA \vdash x\dup\tB\quad \B,x\dup\tB \vdash M\dup\tC}{\B,x\dup\tA \vdash M\dup\tC}
\qquad (C) \quad \db \frac{\B_{1},x\dup\tA \vdash M\dup\tB\quad \B_{2} \vdash N\dup\tA}{\B_{1}, \B_{2} \vdash M[N/x]\dup\tB}
\]
\[(\vee I') \quad \db \frac{\B, x\dup\tA \vdash M\dup\tC\quad \B, x\dup\tB \vdash M\dup\tC}{\B, x\dup\tA\vee\tB \vdash M\dup\tC}
\qquad
(\vee E') \quad \db \frac{\B_{1}, x\dup\tA \vdash M\dup\tC\quad \B_{1}, x\dup\tB \vdash M\dup\tC\quad \B_{2} \vdash N\dup\tA\vee\tB}{\B_{1},\B_{2} \vdash M[N/x]\dup\tC}\]

\noindent
To show $(\vee I')$ it is enough to apply rule $(\vee E)$ with $x\dup\tA\vee\tB \vdash x\dup(\tA\vee\tB)\wedge(\tA\vee\tB)$ as third premise.

\medskip

The system of Figure \ref{tr} can be extended to non-linear terms simply by erasing the condition that, in rules $(\to E)$ and $(\vee E)$, the environments need to be disjoint. It is easy to check that this extended system is conservative over the present one. Therefore the types that can be derived for \fhp s are the same in the two systems, so the present study of type isomorphism holds for the extended system too.

\medskip

In order to show Subject Reduction one can follow the classical approach of \cite{praw65} by considering a sequent formulation of the type assignment system and  showing cut elimination. This is done in \cite{barba} for a system which differs from the present one  for being not relevant, having the universal type and  rule $(\vee E')$ instead of $(\vee E)$. It is just routine to modify that proof by taking
as left and right rules for the $\vee$ constructor:
\[(\vee L) \quad \db \frac{\B, x\dup\tA\wedge\tD \vdash M\dup\tC\quad \B, x\dup\tB\wedge\tD \vdash M\dup\tC}{\B, x\dup(\tA\vee\tB)\wedge\tD \vdash M\dup\tC}\qquad(\vee R)\quad \frac{\B\vdash M\dup\tA\wedge\tD}{\B\vdash M\dup(\tA\vee\tB)\wedge\tD}\quad\frac{\B\vdash M\dup\tB\wedge\tD}{\B\vdash M\dup(\tA\vee\tB)\wedge\tD} \]
 Remark that, considering only linear terms, cut elimination corresponds to standard $\beta$-reduction, while for arbitrary terms  parallel reductions are needed; for details see \cite{barba}.
Therefore one can conclude:
\begin{theorem}[SR] \labelx{srtheorem}
If  $\B \vdash M\dup\tA$ and $M\longrightarrow_{\beta}^*N$, then $\B \vdash
N\dup\tA$.
\end{theorem}

The Subject Reduction Theorem allows one to show the following
corollary.

\begin{corollary} \labelx{cor}
\begin{enumerate}
\item\labelx{cor3} If $~\B\vdash \lambda x.M\dup\tA\to\tB$, then $\B,x\dup\tA \vdash M\dup\tB$.
\item\labelx{cor2} If $~\B,x\dup\tA\vee\tC \vdash M\dup\tB$, then $\B,x\dup\tA \vdash M\dup\tB$  and  $\B,x\dup\tC \vdash M\dup\tB$.
\item\labelx{cor1} If $~\B\vdash \lambda x.M\dup\tA\to\tC$ and $~\B\vdash \lambda x.M\dup\tB\to\tD$, then $~\B\vdash \lambda x.M\dup\tA\wedge\tB\to \tC\wedge\tD$ and $~\B\vdash \lambda x.M\dup\tA\vee\tB\to\tC\vee\tD$.
\item\labelx{cor4} If $~\B\vdash \lambda x.M\dup\tA\to\tB$, then $~\B\vdash \lambda x.M\dup\tA\wedge\tC\to\tB\vee\tD$ for all $\tC,\tD$.\end{enumerate}
\end{corollary}\spaziop
\begin{proof}
(\ref{cor3}).
One gets $\B,x\dup\tA\vdash (\lambda y.M[y/x])x\dup\tB$ by rule $(\to E)$ and $\alpha$-renaming. So the Subject Reduction Theorem (Theorem \ref{srtheorem}) concludes the proof.

\noindent
(\ref{cor2}).
One gets $\B\vdash \lambda y.M[y/x]\dup\tA\vee\tC\to\tB$ by rule $(\to I)$ and $\alpha$-renaming, and $\B,x\dup\tA \vdash (\lambda y.M[y/x])x\dup\tB$ by rules  $(\vee I)$ and $(\to E)$. So Theorem \ref{srtheorem} concludes the proof.

\noindent
(\ref{cor1}).
By Point (\ref{cor3}) $\B,x\dup\tA \vdash M\dup\tC$ and $\B,x\dup\tB \vdash M\dup\tD$, so by rules $(\vee I)$ and $(\vee I')$ one gets $\B,x\dup\tA\vee\tB \vdash M\dup\tC\vee\tD$, which implies the result by rule $(\to I)$. The proof of $~\B\vdash \lambda x.M\dup\tA\wedge\tB\to \tC\wedge\tD$ is similar.

\noindent
(\ref{cor4}).
Obvious because, by Point (\ref{cor3}), $\B, x\dup\tA \vdash M\dup\tB$.
\end{proof}

Also subject expansion holds.
\begin{theorem}[SE] \labelx{setheorem}
If  $M\longrightarrow_{\beta}^*N$ and $\B \vdash
N\dup\tA$, then $\B \vdash M\dup\tA$.
\end{theorem}\spaziop
\begin{proof}
It is enough to show: $\B \vdash M[N/x]\dup\tA$ implies  $\B \vdash (\lambda x.M)N\dup\tA$. The proof is by induction on the derivation of $\B \vdash M[N/x]\dup\tA$.
The only interesting case is when the last applied rule is
\[(\vee E) \quad \db \frac{\B_{1}, x\dup\tC\wedge\tD
\vdash M\dup\sigma\quad \B_{1}, x\dup\tB\wedge\tD \vdash M\dup\sigma\quad
\B_{2} \vdash N\dup(\tC\vee\tB)\wedge\tD}{\B_{1},\B_{2} \vdash
M[N/x]\dup\sigma}\]
It is easy to derive $x\dup(\tC\vee\tB)\wedge\tD \vdash x\dup(\tC\wedge\tD)\vee(\tB\wedge\tD)$. Rule $(\vee I')$ applied to the first two premises gives $\B_{1}, x\dup(\tC\wedge\tD)\vee(\tB\wedge\tD)\vdash M\dup\sigma$. So rule $(L)$ derives $\B_{1}, x\dup(\tC\vee\tB)\wedge\tD\vdash M\dup\sigma$, and rule $(\to I)$ derives $\B_{1}\vdash \lambda x. M\dup (\tC\vee\tB)\wedge\tD\to\sigma$. Rule $(\to E)$ gives the conclusion.
\end{proof}

The following basic isomorphisms are directly related to standard properties of functional types and to set theoretic properties of union and intersection.
It is interesting to remark that all these isomorphisms are provable equalities in the system {\bf B$_+$} of relevant logic \cite{SE3}.

\begin{lemma}\labelx{arrowIso} The following isomorphisms hold:
\[\begin{array}{ll}
 \mbox{\rm{\bfseries idem}.}&  \tA \wedge \tA \approx \tA,\ \tA \vee \tA \approx \tA\\
     \mbox{\rm{\bfseries comm}.}&  \tA \wedge \tB \approx \tB\wedge \tA,\ \tA \vee \tB \approx \tB\vee \tA\\
     \mbox{\rm{\bfseries assoc}.}&  (\tA \wedge \tB)\wedge \tC \approx \tA\wedge (\tB\wedge \tC),\ (\tA \vee \tB)\vee \tC \approx \tA\vee (\tB\vee \tC)\\
   \mbox{\rm{\bfseries dist$\to\wedge$}.}& \tA  \to \tB\wedge \tC \approx (\tA \to \tB)  \wedge (\tA \to \tC) \\
    \mbox{\rm{\bfseries dist$\to\vee$}.}& \tA\vee \tB \to \tC \approx (\tA \to \tC)  \wedge (\tB \to \tC)\\
    \mbox{\rm{\bfseries swap}.}& \tA\to \tB \to \tC \approx \tB \to \tA  \to  \tC\\
  \mbox{\rm{\bfseries dist$\wedge\vee$}.}&   (\tA\vee \tB) \wedge \tC  \approx (\tA\wedge \tC)\vee (\tB \wedge \tC) \\
\mbox{\rm{\bfseries dist$\vee\wedge$}.}& (\tA\wedge\tB)\vee\tC \approx (\tA\vee\tC)\wedge(\tB\vee\tC) \\
\end{array} \]
\end{lemma}\spaziop
\begin{proof}
The $\eta$-expansion of the identity $\lambda xy.xy$ proves the fourth and the fifth isomorphisms, $\lambda xy_{1}y_{2}.xy_{2} y_{1}$ proves the sixth one and the identity proves all the remaining ones.
\end{proof}

The isomorphisms \rm{\bfseries idem}, \rm{\bfseries comm} and \rm{\bfseries assoc} allow one to consider types, at top level, modulo idempotence, commutativity and associativity of $\wedge$ and $\vee$.
Then types, at top level, can be written as $\bigwedge_{i \in I}  \sigma_{i}$   and $\bigvee_{i \in I}  \sigma_{i}$   with finite $I$, where a single arrow or atomic type is seen both as an intersection and as a union (in this case $I$ is a singleton). However, as noted in the introduction, these isomorphisms are not preserved by arbitrary contexts since, for example, $\tA \vee \tB\to \tC \approx \tB \vee \tA \to \tC$ but $(\tA \vee \tB\to \tC) \wedge \varphi$   and  $(\tB \vee \tA \to \tC)\wedge \varphi$ are not isomorphic.

\medskip

The isomorphisms of Lemma \ref{arrowIso} naturally induce the notions of disjunctive and conjunctive forms.  In particular:
\begin{itemize}
\item  the {\em disjunctive weak normal form} of a type $\tA$ (notation \w\tA) is obtained by using \rm{(\bfseries dist$\wedge\vee$)}  from left to right  at top level;
\item the {\em conjunctive weak normal form} of a type $\tA$ (notation \cw\tA) is obtained by using  \rm{(\bfseries dist$\to\wedge$}), \rm{(\bfseries dist$\to\vee$}), and \rm{(\bfseries dist$\vee\wedge$)} from left to right at top level.
\end{itemize}
Notice that the isomorphisms \rm{(\bfseries dist$\to\wedge$}) and \rm{(\bfseries dist$\to\vee$}) are useful only to get conjunctive normal forms, since they only generate intersections.

\medskip

\noindent
This section ends with some lemmas on derivability properties.
Lemmas \ref{var-un-in} characterises the types derivable for variables using disjunctive weak normal form, Corollary \ref{simpleM1} considers three useful particular cases of previous lemma, while Lemma \ref{simpleM2} gives typing properties of the application of a variable to $n$ $\lambda$-terms.

\begin{lemma}\labelx{var-un-in}
If $\w\tA=\bigvee_{i\in I}(\bigwedge_{h\in H_i}\tI^{(i)}_h)$, $\w\tB=\bigvee_{j\in J}(\bigwedge_{k\in K_j}\tL^{(j)}_k)$ and $x\dup\tA\vdash  x\dup\tB$, then for all $i\in I$ there is $j_i\in J$ such that $\set{\tL_k^{(j_i)}\mid k \in K_{j_i}} \subseteq \set{\tI_h^{(i)}\mid h \in H_i}$, which implies $ x\dup\bigwedge_{h\in H_i}\tI^{(i)}_h\vdash  x\dup\bigwedge_{k\in K_{j_i}}\tL^{(j_i)}_k$.
\end{lemma}\spaziop
\begin{proof} By induction on derivations.
Assume $\w\tC=\bigvee_{l\in L}(\bigwedge_{w\in W_l}\tQ^{(l)}_w)$ and $\w\tD=\bigvee_{r\in R}(\bigwedge_{s\in S_r}\tY^{(r)}_s)$ and $\w\tE=\bigvee_{u\in U}(\bigwedge_{v\in V_u}\tV^{(u)}_v)$. If the last applied rule is $(Ax)$ or $(\vee I)$ it is easy.

If the deduction ends with $(\wedge I)$:
\[(\wedge I) \quad  \db \frac{x\dup\tA \vdash x\dup\tC ~~ x\dup\tA \vdash x\dup\tD}
{x\dup\tA \vdash x\dup\tC \wedge \tD}\]
by definition $\w{\tC \wedge \tD}=\bigvee_{l\in L}\bigvee_{r\in R}((\bigwedge_{w\in W_l}\tQ^{(l)}_w)\wedge(\bigwedge_{s\in S_r}\tY^{(r)}_s))$. By induction
for all $i\in I$ there is $l_i\in L$ such that $\set{\tQ_w^{(l_i)}\mid w \in W_{l_i}} \subseteq \set{\tI_h^{(i)}\mid h \in H_i}$ and for all $i\in I$ there is $r_i\in R$ such that $\set{\tY_s^{(r_i)}\mid s \in S_{r_i}} \subseteq \set{\tI_h^{(i)}\mid h \in H_i}$, then  for all $i\in I$  there are $l_i\in L$ and $r_i\in R$ such that $\set{\tQ_w^{(l_i)}\mid w \in W_{l_i}}\cup\set{\tY_s^{(r_i)}\mid s \in S_{r_i}} \subseteq \set{\tI_h^{(i)}\mid h \in H_i}$.

If the deduction ends with $(\wedge E)$:
\[(\wedge E) \quad \db \frac{x\dup\tA \vdash x\dup\tB \wedge \tC}{x\dup\tA \vdash x\dup\tB}\]
by definition  $\w{\tT\wedge\tR$} = $\bigvee_{j\in J}\bigvee_{l\in L}((\bigwedge_{k\in K_j}\tL^{(j)}_k)\wedge(\bigwedge_{w\in W_l}\tQ^{(l)}_w))$. By induction for all $i\in I$ there are  $j_i\in J$ and $l_i\in L$ such that
 $\set{\tL_k^{(j_i)} \mid k \in K_{j_i}} \cup \set{\tQ_w^{(l_i)} \mid w \in W_{l_i}}\subseteq \set{\tI_h^{(i)}\mid h \in H_i}$.

If the deduction ends with $(\vee E)$:
\[(\vee E) \quad \db \frac{y\dup\tC\wedge\tD \vdash y\dup\tB\quad y\dup\tE\wedge\tD \vdash y\dup\tB\quad x\dup\tA \vdash x\dup(\tC\vee\tE)\wedge\tD}{x\dup\tA \vdash x\dup\tB}\]
 By definition $\w{\tC \wedge \tD}=\bigvee_{l\in L}\bigvee_{r\in R}((\bigwedge_{w\in W_l}\tQ^{(l)}_w)\wedge(\bigwedge_{s\in S_r}\tY^{(r)}_s))$ and\\ $\w{\tE \wedge \tD}=\bigvee_{u\in U}\bigvee_{r\in R}((\bigwedge_{v\in V_u}\tV^{(u)}_v)\wedge(\bigwedge_{s\in S_r}\tY^{(r)}_s))$ and $\w{(\tC\vee\tE)\wedge\tD} =$\\\centerline{$ \w{(\tC\wedge\tD)\vee(\tE\wedge\tD)} =
 (\bigvee_{l\in L}\bigvee_{r\in R}((\bigwedge_{w\in W_l}\tQ^{(l)}_w)\wedge(\bigwedge_{s\in S_r}\tY^{(r)}_s)))\vee(\bigvee_{u\in U}\bigvee_{r\in R}((\bigwedge_{v\in V_u}\tV^{(u)}_v)\wedge(\bigwedge_{s\in S_r}\tY^{(r)}_s)))$.} By induction:
\begin{itemize}
\item  on the first premise for all $l\in L$ and $r\in R$ there is $j_{l,r}\in J$ such that\\ $\set{\tL_k^{(j_{l,r})}\mid k \in K_{j_{l,r}}} \subseteq \set{\tQ_w^{(l)}\mid w \in W_l}\cup\set{\tY_s^{(r)}\mid s \in S_r} $ and
\item  on the second premise  for all $u\in U$ and $r\in R$  there is $j_{u,r}\in J$ such that\\ $\set{\tL_k^{(j_{u,r})}\mid k \in K_{j_{u,r}}} \subseteq \set{\tV^{(u)}_v\mid v \in V_u}\cup\set{\tY_s^{(r)}\mid s \in S_r} $ and
\item  on the third premise for all $i\in I$ either there are $l_i\in L$ and $r_i\in R$ such that\\ $\set{\tQ_w^{(l_i)}\mid w \in W_{l_i}}\cup\set{\tY_s^{(r_i)}\mid s \in S_{r_i}}\subseteq \set{\tI_h^{(i)}\mid h \in H_i} $ or  there are $u_i\in U$ and $r_i\in R$ such that\\ $\set{\tV^{(u_i)}_v\mid v \in V_{u_i}}\cup\set{\tY_s^{(r_i)}\mid s \in S_{r_i}}\subseteq \set{\tI_h^{(i)}\mid h \in H_i} $.
\end{itemize}
So for all $i\in I$ there is $j_i\in J$ such that $\set{\tL_k^{(j_i)}\mid k \in K_{j_i}} \subseteq \set{\tI_h^{(i)}\mid h \in H_i}$.
\end{proof}

\begin{corollary} \labelx{simpleM1}
\begin{enumerate}
\item \labelx{ss3}
If $ x\dup\tA \to\tB \vdash x\dup \tC \to\tD$, then $\tA \to \tB = \tC \to \tD$.
\item \labelx{ss1}
If $ x\dup\tA \to\tB \vdash x\dup(\tC\vee\tD)\wedge\tE$, then either $ x\dup\tA \to\tB \vdash x\dup\tC\wedge\tE$ or $ x\dup\tA \to\tB \vdash x\dup\tD\wedge\tE$.
\item \labelx{ss2} 
Let $\ti$ be a union of atomic  and arrow types pairwise different. Then $ x\dup\ti \vdash x\dup\tk$ implies either $\tk = \ti$ or $\tk=\ti\vee\tx$ for some type $\tx$.
\end{enumerate}
\end{corollary}\spaziop
\begin{proof}
(\ref{ss3}),(\ref{ss2}). By Lemma \ref{var-un-in}.\\
(\ref{ss1}). By rule $(\wedge\, E)$, Lemma \ref{var-un-in}  and rule $(\wedge\, I)$. 
\end{proof}
\noindent
Note that Point (\ref{ss2}) of previous corollary holds only under the given condition on type $\ti$, since for example $x\dup(\varphi\to\varphi)\vee(\varphi\to\varphi)\vdash x\dup\varphi\to\varphi$.

\medskip

In the following, as usual, $\rv{\Gamma}{M}$  denotes the set of premises in $\Gamma$ whose subjects are the free variable of $M$.
\begin{lemma} \labelx{simpleM2}
Let $\Gamma_x = \Gamma, x\dup\tB_1\to\ldots \to \tB_n\to \tA$ and
$\Gamma_x \vdash xM_1\ldots M_{n}\dup\tC$. Then:
\begin{enumerate}
\item\labelx{simple1}
 $\Gamma_x  \vdash xM_1\ldots M_{n}\dup\tA$ and $\rv{\Gamma}{M_i}\vdash M_i\dup\tB_i$ for $1 \leq i \leq n$;
\item \labelx{simple2} $y\dup\tA \vdash y\dup\tC$.
\end{enumerate}
\end{lemma}\spaziop
\begin{proof}
A stronger statement is  proved, i.e. that for all types $\tG$:\\
\centerline{$
x\dup\tB_1\to\ldots \to \tB_n\to \tA \vdash x \dup \tG \text{ \;\; and  \;\;}
\Gamma, x\dup \tG \vdash xM_1\ldots M_{n}\dup\tC$}
imply Points (\ref{simple1}) and (\ref{simple2}) above.

\noindent
If $m=0$ Point (\ref{simple1}) is immediate, Point (\ref{simple2}) follows by rule $(L)$. \\
For $m > 0$ the proof is by induction on derivations. 
First note that the last applied rule can be neither $(Ax)$ nor $(\to I)$. If the last applied rule is $(\wedge I)$,  $(\vee I)$ or $(\wedge E)$  Points (\ref{simple1}) and (\ref{simple2}) easily follow.

 If the deduction ends with $(\to E)$:
\[
  (\to\;E) ~~~\frac{\rv{\Gamma}{xM_1\ldots M_{n-1}}, x\dup \tG \vdash xM_1\ldots M_{n-1}\dup\tD \to \tC \quad\rv{\Gamma}{M_{n}} \vdash M_{n}\dup\tD}{\Gamma, x\dup \tG \vdash xM_1\ldots M_{n-1} M_{n}\dup\tC}
\]
By induction, Point (\ref{simple2}) implies $y\dup\tB_n\to\tA\vdash y\dup\tD\to\tC$, which gives $\tD=\tB_n$ and $\tC=\tA$ by Corollary \ref{simpleM1}(\ref{ss3}). This shows Point (\ref{simple2}) and $\rv{\Gamma}{M_{n}} \vdash M_{n}\dup\tB_n$. By induction $\rv{\Gamma}{M_i}\vdash M_i\dup\tB_i$ for $1 \leq i \leq n-1$ and $\rv{\Gamma_x}{xM_1\ldots M_{n-1}}  \vdash xM_1\ldots M_{n-1}\dup\tB_n\to \tA$, so rule $(\to E)$ gives
$\Gamma_x  \vdash xM_1\ldots M_{n}\dup \tA$
and this concludes the proof of Point (\ref{simple1}).

If the deduction ends with $(\vee E)$ two different cases are considered according to the subterms which are the subjects of the premises. In the first case:
\[ (\vee E)~~~\frac{\begin{array}{c}
                    \B_1, z\dup\tD_1\wedge \tF \vdash zM_{s+1}\ldots M_{n}\dup\tC \quad
                    \B_1, z\dup\tD_2\wedge \tF \vdash zM_{s+1}\ldots M_{n}\dup\tC\\
                    \rv{\Gamma}{M_1,\ldots, M_s}, x\dup \tG \vdash xM_1\ldots M_s\dup(\tD_1 \vee \tD_2)\wedge \tF \end{array}}
                     {\Gamma, x\dup \tG \vdash x M_1 \ldots M_{n}\dup\tC}
                        \]
  where $\B_1=\rv{\Gamma}{M_{s+1},\ldots, M_{n}}$ and $0 \leq s \leq n$. By induction the third premise implies\vspace{-8pt}
 \begin{equation}\rv{\Gamma_x}{M_1,\ldots ,M_s} \vdash xM_1 \ldots M_s\dup\tB_{s+1} \to \ldots\to \tB_n\to\tA\labelx{e1}\vspace{-8pt}\end{equation}
  and $\rv{\Gamma}{M_i}\vdash M_i\dup\tB_i$ for $1\leq i\leq s$
  and\vspace{-8pt}
  \begin{equation}
  z\dup\tB_{s+1} \to \ldots\to \tB_n\to\tA \vdash z\dup(\tD_1 \vee \tD_2)\wedge \tF.\labelx{a01}\vspace{-8pt}\end{equation}
  Corollary \ref{simpleM1}(\ref{ss1}) applied  to (\ref{a01}) gives\vspace{-8pt} \begin{equation}z\dup\tB_{s+1} \to \ldots\to \tB_n\to\tA \vdash z\dup\tD_i\wedge \tF\labelx{a02}\vspace{-8pt}\end{equation} where either $i=1$ or $i=2$. Let $i=1$, then
  induction applied to (\ref{a02}) and to the first premise gives\vspace{-8pt}
  \begin{equation}\rv{\Gamma}{M_{s+1}\ldots M_{n}}, z\dup\tB_{s+1} \to \ldots\to \tB_n\to\tA \vdash z M_{s+1}\ldots M_{n}\dup \tA\labelx{e2}\vspace{-8pt}\end{equation}
 and $\rv{\Gamma}{M_i}\vdash M_i\dup\tB_i$ for $s+1\leq i\leq n$  and $y\dup\tA \vdash y\dup\tC$ (i.e. Point (\ref{simple2})).\\
  Rule $(C)$ applied to (\ref{e2}) and (\ref{e1}) derives\\ \centerline{$\Gamma_x \vdash x M_1 \ldots M_{n}\dup\tB_{m+1}\to\ldots \to \tB_n\to \tA$} and this completes the proof of  Point (\ref{simple1}).

  In the second case:
\[ (\vee E)~~~\frac{\begin{array}{c}
                    \B_1, z\dup\tD_1\wedge \tF \vdash xM_1\ldots M_{s-1}MM_{s+1}\ldots M_{n}\dup\tC \qquad
                    \B_1, z\dup\tD_2\wedge \tF \vdash xM_1\ldots M_{s-1}MM_{s+1}\ldots M_{n}\dup \tC \\
                    \rv{\Gamma}{N} \vdash N\dup(\tD_1 \vee \tD_2)\wedge \tF \end{array}}
                     {\Gamma \vdash x M_1 \ldots M_{n}\dup\tC}
                        \]
where $\B_1=\rv{\Gamma}{M_1,\ldots, M_{s-1}MM_{s+1}\ldots M_{n}},x\dup\tG$ and $M_s=M[N/z]$. Induction on the first two premises gives:\vspace{-6pt}
 \begin{equation} \rv{\Gamma}{M}, z\dup\tD_1\wedge \tF  \vdash M\dup\tB_s\qquad\rv{\Gamma}{M}, z\dup\tD_2\wedge \tF  \vdash M\dup\tB_s\labelx{b3}\vspace{-6pt}\end{equation}
 so the application of rule $(\vee E)$ to (\ref{b3}) and to the third premise derives  $\rv{\Gamma}{M_s} \vdash M_s\dup\tB_s$. The other Points follow by induction.
  \end{proof}

\section{Types of finite hereditary permutators}\labelx{AC}\spaziot
Aim of this section is to characterise the types derivable for the \fhp s.
In particular, for an arbitrary \fhp\ $P$ such that $\vdash P\dup\tA\to\tB$,  two crucial properties of the disjunctive and conjunctive weak normal forms of $\tA$  and  $\tB$ are proved:
\begin{itemize}
\item[{\bf P1}]  if $\w\tA=\bigvee_{i\in I}\tM_i$ and $\w\tB=\bigvee_{j\in J}\tN_j$, then for all $i\in I$ there is $j_i\in J$ such that  \mbox{$\vdash P\dup\tM_i\to\tN_{j_i}$;}
\item[{\bf P2}]  if $\cw\tA=\bigwedge_{i\in I}\ti_i$ and  $\cw\tB=\bigwedge_{j\in J}\tk_j$, then for all $j\in J$ there is $i_j\in I$ such that  \mbox{$\vdash P\dup\ti_{i_j}\to\tk_{j}$.}
\end{itemize}

\noindent The content of this section can be summarised as follows:
\begin{itemize}
\item Lemma \ref{pfhp-sub}  gives all possible ways of getting an \fhp , possibly with some missing abstractions, as result of a substitution (this is useful to deal with rule $(\vee E)$);
\item Theorem \ref{dis} characterises the types derivable for  \fhp s using disjunctive weak normal form: it gives {\bf P1};
\item Lemma \ref{con} characterises the types derivable for \fhp s, possibly with some missing abstractions, using conjunctive weak normal form: it implies {\bf P2}, i.e. Theorem \ref{last}.
\end{itemize}
The proof of {\bf P1} is much simpler than that of {\bf P2}. The reason is that Theorem \ref{dis} uses the property of union stated in Corollary \ref{cor}(\ref{cor2}), while there is no similar property for intersection.

\begin{lemma}\labelx{pfhp-sub}
If $\lambda x y_{1}\ldots y_{n}.xQ_1\ldots Q_n$ $(n \geq  0)$ is an \fhp\ and $\lambda y_{m+1}\ldots y_{n}.xQ_1\ldots Q_n=M[N/z]$ with $0 \leq m\leq n$\footnote{for $m=n$ 
$M[N/z]=xQ_1\ldots Q_n$.}, then the possible cases are:
\begin{enumerate}
\item\labelx{pfhp-sub1} $M= \lambda y_{m+1}\ldots y_{n}.zQ_{l+1}\ldots Q_n$ with $l\leq m$ and $N=  xQ_1\ldots Q_l$ and $FV(N)\subseteq\set{x,y_{1},\ldots, y_{m}}$;
\item\labelx{pfhp-sub2} $M=  z$ and $N=  \lambda y_{m+1}\ldots y_{n}.xQ_1\ldots Q_n$;
\item\labelx{pfhp-sub3} $M= \lambda y_{m+1}\ldots y_{n}.xQ_1\ldots Q_{j-1}zQ_{j+1}\ldots Q_n$ and $N=  Q_j$ and the head variable of $Q_j$ belongs to $\set{y_{1},\ldots, y_{m}}$;
\item\labelx{pfhp-sub4} $M= \lambda y_{m+1}\ldots y_{n}.xQ_1\ldots Q_{j-1}Q'_jQ_{j+1}\ldots Q_n$ and $Q'_j=Q_j[z/y_l]$ and $N=y_l$, where $y_l\in\set{y_{1},\ldots, y_{m}}$ is the head variable of $Q_j$.
\end{enumerate}
\end{lemma}\spaziop
\begin{proof}
Easy observing that the variables $y_1,..., y_n$ must be the head variables of $Q_1,..., Q_n$.
\end{proof}

\begin{theorem}[Property {\bf P1}]\labelx{dis}
Let $\w\tA=\bigvee_{i\in I}\tM_i$, $\w\tB=\bigvee_{j\in J}\tN_j$ and $P$ be an \fhp. Then $\vdash P\dup\tA\to\tB$ implies that for all $i\in I$ there is $j_i\in J$ such that  $\vdash P\dup\tM_i\to\tN_{j_i}$.
\end{theorem}\spaziop
\begin{proof}If $P=\lambda x.x$ the proof follows immediately from Lemma \ref{var-un-in}.

 Otherwise let $P=\lambda x y_{1}\ldots y_{n}.xQ_1\ldots Q_n$. By Corollary \ref{cor}(\ref{cor3}) $x\dup\tA\vdash\lambda y_{1}\ldots y_{n}.xQ_1\ldots Q_n\dup\tB$. The proof is
by induction on the derivation of $x\dup\tA\vdash\lambda y_{1}\ldots y_{n}.xQ_1\ldots Q_n\dup\tB$. Assume $\w\tC=\bigvee_{h\in H}\tl_h$ and $\w\tD=\bigvee_{k\in K}\te_k$ and $\w\tE=\bigvee_{l\in L}\tv_l$. Let $Q=\lambda y_{1}\ldots y_{n}.xQ_1\ldots Q_n$.

If the last applied rule is $(\vee I)$ the proof is easy.

If the last applied rule is $(\to I)$, then $\tB$ is an arrow type.  Corollary \ref{cor}(\ref{cor2}) gives  $x\dup\tM_i\vdash Q\dup\tB$ for all $i\in I$.

Let the last applied rule be $(\wedge E)$:
\[(\wedge E) \quad  \db \frac{x\dup\tA \vdash Q\dup\tB \wedge \tC}{x\dup\tA \vdash Q\dup\tB}
\]
By definition $\w{\tB \wedge \tC}=\bigvee_{j\in J}\bigvee_{h\in H}(\tN_j\wedge\tl_h)$. By induction,
for all $i\in I$ there are $j_i\in J$  and $h_i\in H$ such that $x\dup\tM_i\vdash Q\dup\tN_{j_i}\wedge\tl_{h_i}$, which gives $x\dup\tM_i\vdash Q\dup\tN_{j_i}$ for all $i\in I$ using rule $(\wedge E)$.

Let the last applied rule be $(\wedge I)$:
\[(\wedge I) \quad  \db \frac{x\dup\tA \vdash Q\dup\tC ~~ x\dup\tA \vdash Q\dup\tD}
{x\dup\tA \vdash Q\dup\tC \wedge \tD}\]
By definition $\w{\tC \wedge \tD}=\bigvee_{h\in H}\bigvee_{k\in K}(\tl_h\wedge\te_k)$.\\ By induction,
for all $i\in I$ there is $h_i\in H$ such that $x\dup\tM_i\vdash Q\dup\tl_{h_i}$ and for all $i\in I$ there is $k_i\in K$ such that $x\dup\tM_i\vdash Q\dup\te_{k_i}$. Then rule $(\wedge I)$ derives $x\dup\tM_i\vdash Q\dup\tl_{h_i}\wedge\te_{k_i}$ for all $i\in I$.

If the last applied rule is $(\vee E)$ by Lemma \ref{pfhp-sub} there are two cases to consider.\\ By definition, $\w{\tC \wedge \tD}=\bigvee_{h\in H}\bigvee_{k\in K}(\tl_h\wedge\te_k)$ and $\w{\tE \wedge \tD}=\bigvee_{l\in L}\bigvee_{k\in K}(\tv_l\wedge\te_k)$ and $\w{(\tC\vee\tE)\wedge\tD}=(\bigvee_{h\in H}\bigvee_{k\in K}(\tl_h\wedge\te_k))\vee(\bigvee_{l\in L}\bigvee_{k\in K}(\tv_l\wedge\te_k))$. In the first case:
\[(\vee E) \quad \db \frac{ z\dup\tC\wedge\tD \vdash Q'\dup\tB\quad z\dup\tE\wedge\tD \vdash Q'\dup\tB\quad x\dup\tA \vdash x\dup(\tC\vee\tE)\wedge\tD}{x\dup\tA \vdash Q\dup\tB}\]
where $Q'=\lambda y_{1}\ldots y_{n}.zQ_1\ldots Q_n$. By induction:
\begin{itemize}
\item  on the first premise for all $h\in H$ and $k\in K$ there is $j_{h,k}\in J$ such that $z\dup\tl_h\wedge\te_k\vdash Q'\dup\tN_{j_{h,k}}$;
\item  on the second premise  for all $l\in L$ and $k\in K$ there is $j_{l,k}\in J$ such that $z\dup\tv_l\wedge\te_k\vdash Q'\dup\tN_{j_{l,k}}$;
\item on the third premise for all $i\in I$ either there are $h_i\in H$ and $k_i\in K$ such that $x\dup\tM_i\vdash x\dup\tl_{h_i}\wedge\te_{k_i}$ or  there are $l_i\in L$ and $k_i\in K$ such that $x\dup\tM_i\vdash x\dup\tv_{l_i}\wedge\te_{k_i}$.
\end{itemize}
So rule $(L)$ implies that for all $i\in I$ either there is $j_{h_i,k_i}\in J$ such that $x\dup\tM_i\vdash Q\dup\tN_{j_{h_i,k_i}}$ or there is $j_{l_i,k_i}\in J$ such that $x\dup\tM_i\vdash Q\dup \tN_{j_{l_i,k_i}}$. The other possible case is given by:
\[(\vee E) \quad \db \frac{z\dup\tC\wedge\tD \vdash z\dup\tB\quad z\dup\tE\wedge\tD \vdash z\dup \tB\quad x\dup\tA \vdash Q\dup(\tC\vee\tE)\wedge\tD}{x\dup\tA \vdash Q\dup\tB}\]
By induction:
\begin{itemize}
\item on the first premise  for all $h\in H$ and $k\in K$ there is $j_{h,k}\in J$ such that $z\dup\tl_h\wedge\te_k\vdash z\dup\tN_{j_{h,k}}$;
\item  on the second premise for all $l\in L$ and $k\in K$ there is $j_{l,k}\in J$ such that $z\dup\tv_l\wedge\te_k\vdash z\dup\tN_{j_{l,k}}$;
\item on the third premise for all $i\in I$ either there are $h_i\in H$ and $k_i\in K$ such that $x\dup\tM_i\vdash Q\dup\tl_{h_i}\wedge\te_{k_i}$ or  there are $l_i\in L$ and $k_i\in K$ such that $x\dup\tM_i\vdash Q\dup\tv_{l_i}\wedge\te_{k_i}$.
\end{itemize}
So rule $(C)$ implies that for all $i\in I$ either there is $j_{h_i,k_i}\in J$ such that $x\dup\tM_i\vdash Q\dup \tN_{j_{h_i,k_i}}$ or there is $j_{l_i,k_i}\in J$ such that $x\dup\tM_i\vdash Q\dup\tN_{j_{l_i,k_i}}$.
\end{proof}

\begin{lemma}\labelx{con}
Let $\cw\tA=\bigwedge_{i\in I}\ti_i$, $\cw\tB=\bigwedge_{j\in J}\tk_j$ and $\lambda x y_{1}\ldots y_{n}.xQ_1\ldots Q_n$ $(n \geq  0)$ be an \fhp. Then
 $x\dup \tA, y_1\dup \tC_1,\ldots,y_m\dup \tC_m\vdash \lambda y_{m+1}\ldots y_{n}.xQ_1\ldots Q_n\dup \tB$ and  $\w{\tC_h}=\bigvee_{k\in K_h}\tM^{(h)}_k$ $(1\leq h\leq m\leq n)$ imply that
 for all $j\in J$ and for all $k_h\in K_h$ $(1\leq h\leq m)$ there is $i_{j,k_1,\ldots,k_m}\in I$ such that $$x\dup \ti_{i_{j,k_1,\ldots,k_m}}, y_1\dup \tM^{(1)}_{k_1},\ldots,y_m\dup \tM^{(m)}_{k_m}\vdash \lambda y_{m+1}\ldots y_{n}.xQ_1\ldots Q_n\dup \tk_{j}.$$
\end{lemma}\spaziop
\begin{proof}
By induction on derivations.  If the last applied rule is $(Ax)$ or $(\wedge I)$  the proof is easy.

Assume $\cw\tD=\bigwedge_{l\in L}\tx_l$ and $\cw\tE=\bigwedge_{s\in S}\tr_s$ and $\cw\tF=\bigwedge_{t\in T}\tu_t$. Then $\cw{\tD \wedge \tE}=\bigwedge_{l\in L}\tx_l\wedge\bigwedge_{s\in S}\tr_s$ and $\cw{\tF \wedge \tE}=\bigwedge_{t\in T}\tu_t\wedge\bigwedge_{s\in S}\tr_s$ and $\cw{(\tD\vee\tF)\wedge\tE}=\bigwedge_{l\in L}\bigwedge_{t\in T}(\tx_l\vee\tu_t)\wedge\bigwedge_{s\in S}\tr_s$. %

Assume $\w\tD=\bigvee_{l\in L}\tN_l$ and $\w\tE=\bigvee_{s\in S}\tl_s$ and $\w\tF=\bigvee_{t\in T}\te_t$. Then $\w{\tD \wedge \tE}=\bigvee_{l\in L}\bigvee_{s\in S}(\tN_l\wedge\tl_s)$ and $\w{\tF \wedge \tE}=\bigvee_{t\in T}\bigvee_{s\in S}(\te_t\wedge\tl_s)$ and  $\w{(\tD\vee\tF)\wedge\tE}=\bigvee_{l\in L}\bigvee_{s\in S}(\tN_l\wedge\tl_s)\vee\bigvee_{t\in T}\bigvee_{s\in S}(\te_t\wedge\tl_s)$.

 Let the last applied rule be $(\to I)$:
\[(\to I) \quad  \db \frac{\Gamma,y_{m+1}\dup \tC_{m+1} \vdash R\dup  \tD}
{\Gamma\vdash \lambda y_{m+1}.R\dup  \tC_{m+1} \to \tD}\]
where $\Gamma=x\dup \tA, y_1\dup \tC_1,\ldots,y_m\dup \tC_m$ and $R=\lambda y_{m+2}\ldots y_{n}.xQ_1\ldots Q_n$.\\  By definition  $\cw{\tC_{m+1} \to \tD}=\bigwedge_{l\in L}\bigwedge_{k\in K_{m+1}}(\tM^{(m+1)}_k \to \tx_l)$. By induction, for all $l\in L$ and for all $k_h\in K_h$ $(1\leq h\leq m+1)$ there is $i_{l,k_1,\ldots,k_{m+1}}\in I$ such that  $x\dup \ti_{i_{l,k_1,\ldots,k_{m+1}}}, y_1\dup \tM^{(1)}_{k_1},\ldots,y_{m+1}\dup \tM^{(m+1)}_{k_{m+1}}\vdash R\dup \tx_{l}$, so the application of rule $(\to I)$ concludes the proof.

Let the last applied rule be $(\to E)$:
\[(\to E) \quad  \db \frac{\Gamma \vdash R\dup \tD\to\tB\quad y_{\pi(r)}\dup\tC_{\pi(r)} \vdash Q_{r}\dup \tD}
{\Gamma \vdash RQ_{r}\dup \tB}\]
where $\Gamma=x\dup\tA, y_{\pi(1)}\dup\tC_{\pi(1)},\ldots,y_{\pi(r-1)}\dup\tC_{\pi(r-1)}$ and  $R=xQ_1\ldots Q_{r-1}$. If $\w\tD=\bigvee_{u\in U}\tN_u$, then  $\cw{\tD \to \tB}=\bigwedge_{u\in U}\bigwedge_{j\in J}(\tN_u \to \tk_j)$. By induction for all $u\in U$, $j\in J$ and for all $k_{\pi(s)}\in K_{\pi(s)}$ $(1\leq s\leq r-1)$ there is $i_{(u,j),k_{\pi(1)},\ldots,k_{\pi(r-1)}}\in I$ such that \vspace{-8pt}
\begin{equation}x\dup\ti_{i_{(u,j),k_{\pi(1)},\ldots,k_{\pi(r-1)}}}, y_1\dup\tM^{({\pi(1)})}_{k_{\pi(1)}},\ldots,y_{\pi(r-1)}\dup\tM^{({\pi(r-1)})}_{k_{\pi(r)}}\vdash R\dup\tN_u\to \tk_j. \vspace{-8pt}\labelx{a3}\end{equation}
 By Theorem \ref{dis} for all $k_{\pi(r)}\in K_{\pi(r)}$ there is $u_{k_{\pi(r)}}\in U$ such that \vspace{-8pt}
  \begin{equation}y_{\pi(r)}\dup\tM^{(\pi(r))}_{k_{\pi(r)}} \vdash Q_{r}\dup \tN_{u_{k_{\pi(r)}}}. \vspace{-8pt}\labelx{a4}\end{equation}
Choosing $u=u_{k_{\pi(r)}}$ in (\ref{a3}) the application of rule $(\to E)$ to (\ref{a3}) and (\ref{a4}) gives the result.

Let the last applied rule be $(\wedge E)$:
\[(\wedge E) \quad \db \frac{\Gamma \vdash R\dup  \tB \wedge \tD}{\Gamma \vdash R\dup \tB}\]
where 
$\Gamma=x\dup \tA, y_1\dup \tC_1,\ldots,y_m\dup \tC_m$ and $R=\lambda y_{m+1}\ldots y_{n}.xQ_1\ldots Q_n$. Since $\cw{\tB\wedge \tD}= \cw\tB\wedge\cw \tD$, this case easily follows by induction.

Let the last applied rule be $(\vee I)$:
\[(\vee I) \quad \db \frac{\Gamma \vdash R\dup  \tD }{\Gamma \vdash R\dup \tD\vee\tE}\]
where
$\Gamma=x\dup \tA, y_1\dup \tC_1,\ldots,y_m\dup \tC_m$ and $R=\lambda y_{m+1}\ldots y_{n}.xQ_1\ldots Q_n$. Since $\cw{\tD\vee\tE}= \bigwedge_{l\in L}\bigwedge_{s\in S}(\tx_l\vee\tr_s)$, this case easily follows by induction.

If the last applied rule is $(\vee E)$ there are four different cases as prescribed by Lemma \ref{pfhp-sub}.\\ In the first case:

\[(\vee E) \quad \db \frac{\Gamma_1, z\dup \tD\wedge\tE \vdash R\dup  \tB\quad ,\Gamma_1, z\dup \tF\wedge\tE \vdash R\dup  \tB\quad \Gamma_2 \vdash xQ_1\ldots Q_u\dup  (\tD\vee\tF)\wedge\tE}{\Gamma_1,\Gamma_2 \vdash \lambda y_{m+1}\ldots y_{n}.xQ_{1}\ldots Q_n\dup \tB}\]
where $R=\lambda y_{m+1}\ldots y_{n}.zQ_{u+1}\ldots Q_n$, $\Gamma_1 =\rv{\set{y_1\dup \tC_1,\ldots,y_m\dup \tC_m}}{R}$,\\  $\Gamma_2 = x\dup \tA,y_{\pi(1)}\dup \tC_{\pi(1)},\ldots,y_{\pi(u)}\dup \tC_{\pi(u)}$ and $(0\leq u\leq m)$. Let $\Gamma_1 = y_{w_1}\dup \tC_{w_1},\ldots,y_{w_{m-u}}\dup \tC_{w_{m-u}}$.

.

For all $j\in J$ and for all $k_{w_v}\in K_{w_v}$ $(1\leq v\leq m-u)$ by induction :
\begin{itemize}
\item  on the first premise  either there is $l_{j,k_{w_1},\ldots,k_{w_{m-u}}}\in L$ such that \\
$z\dup \tx_{l_{j,k_{w_1},\ldots,k_{w_{m-u}}}},  y_{w_1}\dup \tM^{(w_1)}_{k_{w_1}},\ldots,y_{w_{m-u}}\dup \tM^{(w_{m-u})}_{k_{w_{m-u}}}\vdash R\dup \tk_{j}$ or  there is\\ $s_{j,k_{w_1},\ldots,k_{w_{m-u}}}\in S$ such that $z\dup \tr_{s_{j,k_{w_1},\ldots,k_{w_{m-u}}}},  y_{w_1}\dup \tM^{(w_1)}_{k_{w_1}},\ldots,y_{w_{m-u}}\dup \tM^{(w_{m-u})}_{k_{w_{m-u}}}\vdash R\dup \tk_{j}$;
\item  on the second premise  either there is $t_{j,k_{u+1},\ldots,k_m}\in T$ such that \\ $z\dup \tu_{t_{j,k_{w_1},\ldots,k_{w_{m-u}}}},  y_{w_1}\dup \tM^{(w_1)}_{k_{w_1}},\ldots,y_{w_{m-u}}\dup \tM^{(w_{m-u})}_{k_{w_{m-u}}}\vdash R\dup \tk_{j}$ or there is\\ $s_{j,k_{w_1},\ldots,k_{w_{m-u}}}\in S$ such that $z\dup \tr_{s_{j,k_{w_1},\ldots,k_{w_{m-u}}}},  y_{w_1}\dup \tM^{(w_1)}_{k_{w_1}},\ldots,y_{w_{m-u}}\dup \tM^{(w_{m-u})}_{k_{w_{m-u}}}\vdash R\dup \tk_{j}$.
\end{itemize}
Therefore for all $j\in J$ and for all $k_{w_v}\in K_{w_v}$ $(1\leq v\leq m-u)$:
\begin{itemize}
\item either there is $l_{j,k_{w_1},\ldots,k_{w_{m-u}}}\in L$ such that
\vspace{-8pt}\begin{equation}z\dup \tx_{l_{j,k_{w_1},\ldots,k_{w_{m-u}}}},  y_{w_1}\dup \tM^{(w_1)}_{k_{w_1}},\ldots,y_{w_{m-u}}\dup \tM^{(w_{m-u})}_{k_{w_{m-u}}}\vdash R\dup \tk_{j}\vspace{-8pt}\labelx{a5}\end{equation}
 and  there is $t_{j,k_{w_1},\ldots,k_{w_{m-u}}}\in T$ such that \vspace{-8pt}
\begin{equation}z\dup \tu_{t_{j,k_{w_1},\ldots,k_{w_{m-u}}}},  y_{w_1}\dup \tM^{(w_1)}_{k_{w_1}},\ldots,y_{w_{m-u}}\dup \tM^{(w_{m-u})}_{k_{w_{m-u}}}\vdash R\dup \tk_{j};\labelx{a6}\vspace{-8pt}\end{equation}
\item  or there is $s_{j,k_{u+1},\ldots,k_m}\in S$ such that \vspace{-8pt}
\begin{equation} z\dup \tr_{s_{j,k_{w_1},\ldots,k_{w_{m-u}}}},  y_{w_1}\dup \tM^{(w_1)}_{k_{w_1}},\ldots,y_{w_{m-u}}\dup \tM^{(w_{m-u})}_{k_{w_{m-u}}}\vdash R\dup \tk_{j}.\vspace{-8pt}\labelx{a7}\end{equation}
\end{itemize}
By induction the third premise implies that for all $l\in L$, $t\in T$ and for all $k_{\pi(h)}\in K_{\pi(h)}$ $(1\leq h\leq u)$ there is $i_{l,t,k_{\pi(1)},\ldots,k_{\pi(u)}}\in I$ such that \vspace{-8pt}
\begin{equation} x\dup \ti_{i_{l,t,k_{\pi(1)},\ldots,k_{\pi(u)}}}, y_{\pi(1)}\dup \tM^{(\pi(1))}_{k_\pi(1)},\ldots,y_{\pi(u)}\dup \tM^{(\pi(u))}_{k_{\pi(u)}}\vdash xQ_1\ldots Q_u\dup \tx_l\vee\tu_t\vspace{-8pt}\labelx{a8}\end{equation} and
for all $s\in S$ and for all $k_{\pi(h)}\in K_{\pi(h)}$ $(1\leq h\leq u)$ there is $i_{s,k_{\pi(1)},\ldots,k_{\pi(u)}}\in I$ such that \vspace{-8pt}
\begin{equation} x\dup \ti_{i_{s,k_{\pi(1)},\ldots,k_{\pi(u)}}},  y_{\pi(1)}\dup \tM^{(\pi(1))}_{k_\pi(1)},\ldots,y_{\pi(u)}\dup \tM^{(\pi(u))}_{k_{\pi(u)}}\vdash xQ_1\ldots Q_u\dup \tr_s.\labelx{a9}\vspace{-8pt}\end{equation}
If (\ref{a5}) and (\ref{a6}) hold, then the conclusion follows from the application of rule $(\vee E)$ to (\ref{a5}),  (\ref{a6}) and (\ref{a8}) by choosing $l=l_{j,k_{w_1},\ldots,k_{w_{m-u}}}$ and $t=t_{j,k_{w_1},\ldots,k_{w_{m-u}}}$. Otherwise (\ref{a7}) must hold, and the conclusion follows from the application of rule $(C)$ to (\ref{a7}) and  (\ref{a9}) by choosing $s=s_{j,k_{u+1},\ldots,k_m}$.

In the second case:

\[(\vee E) \quad \db \frac{z\dup\tD\wedge\tE \vdash z\dup \tB\quad z\dup\tF\wedge\tE \vdash z\dup \tB\quad \Gamma \vdash \lambda y_{m+1}\ldots y_{n}.xQ_{1}\ldots Q_n\dup (\tD\vee\tF)\wedge\tE}{\Gamma\vdash \lambda y_{m+1}\ldots y_{n}.xQ_{1}\ldots Q_n\dup\tB}\]
where $\Gamma=x\dup\tA,y_1\dup\tC_1,\ldots,y_m\dup\tC_m$.

For all $j\in J$ by induction:
\begin{itemize}
\item  the first premise implies that either there is $l_{j}\in L$ such that  $z\dup\tx_{l_{j}}\vdash z\dup\tk_{j}$ or  there is $s_{j}\in S$ such that $z\dup\tr_{s_{j}}\vdash z\dup\tk_{j}$;
\item  the second premise implies that either there is $t_{j}\in T$ such that  $z\dup\tu_{t_{j}}\vdash z\dup\tk_{j}$ or there is $s_{j}\in S$ such that $z\dup\tr_{s_{j}}\vdash z\dup\tk_{j}$.
\end{itemize}
Therefore for all $j\in J$:
\begin{itemize}\setlength{\itemsep}{-.6em}
\item either there are $l_{j}\in L$ and $t_{j}\in T$ such that \vspace{-8pt}
\begin{equation}z\dup\tx_{l_{j}}\vdash z\dup\tk_{j}\qquad z\dup\tu_{t_{j}}\vdash z\dup\tk_{j};\vspace{-8pt}\labelx{a10}\end{equation}
\item  or there is $s_{j}\in S$ such that\vspace{-8pt}
\begin{equation} z\dup\tr_{s_{j}}\vdash z\dup\tk_{j}.\vspace{-8pt}\labelx{a11}\end{equation}
\end{itemize}
By induction on the third premise  for all $l\in L$, $t\in T$ and for all $k_h\in K_h$ $(1\leq h\leq m)$ there is $i_{l,k_1,\ldots,k_m}\in I$ such that\vspace{-4pt}
\begin{equation} x\dup\ti_{i_{l,k_1,\ldots,k_m}}, y_1\dup\tM^{(1)}_{k_1},\ldots,y_m\dup\tM^{(m)}_{k_m}\vdash \lambda y_{m+1}\ldots y_{n}.xQ_1\ldots Q_n\dup\tx_l\vee\tu_t\vspace{-4pt}\labelx{a12}\end{equation} and
for all $s\in S$ and $k_h\in K_h$ $(1\leq h\leq m)$ there is $i_{s,k_1,\ldots,k_m}\in I$ such that\vspace{-4pt}
\begin{equation} x\dup\ti_{s_{l,k_1,\ldots,k_m}}, y_1\dup\tM^{(1)}_{k_1},\ldots,y_m\dup\tM^{(m)}_{k_m}\vdash \lambda y_{m+1}\ldots y_{n}.xQ_1\ldots Q_n\dup\tr_{s}.\vspace{-4pt}\labelx{a13}\end{equation}
If (\ref{a10}) holds, then the conclusion follows from the application of rule $(\vee E)$ to (\ref{a10}) and (\ref{a12}) by choosing $l=l_{j}$ and $t=t_{j}$. Otherwise (\ref{a11}) must hold, and the conclusion follows from the application of rule $(C)$ to (\ref{a11}) and  (\ref{a13}) by choosing $s=s_{j}$.

As for the third case:
\[(\vee E) \quad \db \frac{\Gamma,z\dup\tD\wedge\tE \vdash R\dup \tB\quad  \Gamma,z\dup\tF\wedge\tE \vdash R\dup \tB\quad y_{u}\dup\tC_{u} \vdash Q_v\dup (\tD\vee\tF)\wedge\tE}{\Gamma\vdash \lambda y_{m+1}\ldots y_{n}.xQ_{1}\ldots  Q_n\dup\tB}\]
where $\Gamma=x\dup\tA, y_1\dup\tC_1,\ldots,y_{u-1}\dup\tC_{u-1},y_{u+1}\dup\tC_{u+1}, \ldots,y_m\dup\tC_m$, $R=\lambda y_{m+1}\ldots y_{n}.xQ_{1}\ldots Q_{v-1}zQ_{v+1}\ldots Q_n$ and $u=\pi(v)$.\\
By induction on the first premise for all $j\in J$, $l\in L$, $s\in S$, and for all $k_h\in K_h$ $(1\leq h\leq m, h\not=u)$,  there is $i_{(j,l,s),k_1,\ldots,k_{u-1},k_{u+1},\ldots, k_{m}}\in I$ such that \vspace{-4pt}
\begin{equation}x\!\dup\!\ti_{i_{(j,l,s),k_1,\ldots,k_{u-1},k_{u+1},\ldots, k_{m}}},\! y_1\!\dup\!\tM^{(1)}_{k_1},\ldots,y_{u-1}\!\dup\!\tM^{(u-1)}_{k_{u-1}},y_{u+1}\!\dup\!\tM^{(u+1)}_{k_{u+1}}, \ldots,y_{m}\!\dup\!\tM^{(m)}_{k_{m}},\! z\dup\tN_l\wedge\tl_s\!\vdash\! R\!\dup\!\tk_j.\vspace{-4pt}\labelx{a14}\end{equation}
By induction on the second premise
for all $j\in J$, $t\in T$, $s\in S$, and for all $k_h\in K_h$ $(1\leq h\leq m+1, h\not=u)$, there is $i_{(j,t,s),k_1,\ldots,k_{u-1},k_{u+1},\ldots, k_m}\in I$ such that \vspace{-4pt}
\begin{equation}x\!\dup\!\ti_{i_{(j,t,s),k_1,\ldots,k_{u-1},k_{u+1},\ldots, k_{m}}}, y_1\dup\tM^{(1)}_{k_1},\ldots,y_{u-1}\!\dup\!\tC_{u-1},y_{u+1}\!\dup\!\tC_{u+1}, \ldots,y_{m}\!\dup\!\tM^{(m)}_{k_{m}}, z\!\dup\!\te_t\wedge\tl_s\vdash R\!\dup\!\tk_j.\labelx{a15}\vspace{-4pt}\end{equation}
By Theorem \ref{dis} applied to the third premise for all $k_u\in K_u$:\vspace{-4pt}
\begin{equation}\text{either there are $l_{k_u}\in L$, $s_{k_u}\in S$ such that $y_u\dup\tM^{(u)}_{k_u}\vdash Q_v\dup\tN_{l_{k_u}}\wedge\tl_{s_{k_u}}$;}\vspace{-8pt}\labelx{a16}\end{equation}\vspace{-4pt}
\begin{equation}\text{or there are $t_{k_u}\in T$, $s_{k_u}\in S$ such that $y_u\dup\tM^{(u)}_{k_u}\vdash Q_v\dup\te_{t_{k_u}}\wedge\tl_{s_{k_u}}$.}
\labelx{a17}\end{equation}
If (\ref{a16}) holds, then the conclusion follows from the application of rule $(C)$ to (\ref{a14}) and (\ref{a16}) by choosing $l=l_{k_u}$ and $s=s_{k_u}$. Otherwise (\ref{a15}) must hold, and the conclusion follows from the application of rule $(C)$ to (\ref{a15}) and  (\ref{a17}) by choosing $t=t_{k_u}$ and $s=s_{k_u}$.

The proof for the last case is similar and simpler than that one of the third case.
\end{proof}

\begin{theorem}[Property {\bf P2}]\labelx{last}
If $P$ is an \fhp, $\cw\tA=\bigwedge_{i\in I}\ti_i$, $\cw\tB=\bigwedge_{j\in J}\tk_j$, and $\vdash P\dup\tA\to\tB$, then for all $j\in J$ there is $i_{j}\in I$ such that $\vdash P\dup\ti_{i_{j}}\to\tk_j$.
\end{theorem}

\section{Normalisation of types}\labelx{tn}\spaziot
    To investigate type isomorphism it is necessary to consider the basic laws introduced in Lemma \ref{arrowIso}, for finding conditions
   allowing to apply them  also at the level of subtypes and to exploit some provable  properties
   of type inclusion. To this aim, following  a common approach \cite{BruceDicosmoLongo92,DDGT10},
      a notion of \emph{normal form} of types is introduced. {\em Normal type} is short for type in normal form.


 Normal types are obtained by applying as far as possible a set of isomorphism preserving transformations,
 that are all realised by suitable $\eta$-expansions of the identity. The transformations applied to obtain normal types are essentially:


 \begin{itemize}
 \item the distribution of intersections over unions or vice versa, in such a way that all types to the right of an arrow
 are in  conjunctive normal form and all types to the left of an arrow are in disjunctive normal form. This is obtained by using  \rm{(\bfseries dist$\vee\wedge$)} and \rm{(\bfseries dist$\wedge\vee$)} (\emph{distribution});


 \item the elimination of intersections to the right of arrows and of unions to the left of arrows using the
 isomorphisms \rm{(\bfseries dist$\to\wedge$)} and \rm{(\bfseries dist$\to\vee$)} from left to right (\emph{splitting});
 \item the elimination of redundant intersections and unions, corresponding roughly to intersections and unions performed on types  provably included in one another, as $(\sigma \rightarrow \tau)\wedge (\sigma\vee\rho \to \tau)$, that can be reduced to  $\sigma\vee\rho \to \tau$; similarly $(\sigma \to \rho \vee\tau)\wedge (\sigma \to \tau)$ can be reduced to $\sigma \to \tau$ (\emph{erasure});
 \item the transformation of types at top level in conjunctive normal form.
\end{itemize}
\noindent
 For example the type  $((\varphi_1 \wedge \varphi_2 \to \varphi_2 \vee \varphi_3)\vee (\varphi_2 \to \varphi_5)) \wedge ((\varphi_2 \wedge\varphi_3 \to \varphi_5)\vee(\varphi_4 \to \varphi_3\vee\varphi_5))$ is normal.

\medskip

The  normalisation process, although rather intuitive, needs some care when performed inside a type context since
the used transformations must be isomorphism preserving.

The following  subsection defines the normalisation rules. In Subsection \ref{AB} the soundness and termination of the normalisation
rules and the unicity of normal forms are proved.
The notion of normal form is effective since an algorithm to find the normal form of an arbitrary type can be given. Lastly Subsection \ref{AD} presents interesting properties of normal types, in particular Theorem \ref{discon} characterises the isomorphic normal types.

\subsection{Normalisation rules}\spazios

Since the normalisation rules have to be applied (whenever possible) also to subtypes, the (standard) notion of type \emph{context} is introduced.
\[\C{~}~::=~[~] ~\mid ~\C{~}\to\tS~\mid~ \tS\to\C{~}~\mid~ \tS\wedge\C{~}~\mid~\C{~}\wedge\tS~\mid~ \tS\vee\C{~}~\mid~ \C{~}\vee\tS.\]
The possibility of applying transformations to subtypes strongly depends on the context in which they occur.
An example of this problem was already given at page \pageref{page2}.  Also
the types  $(\sigma \vee \tau \rightarrow \rho) \wedge (\sigma \rightarrow \rho)$ and  $\sigma\vee\tau \to \rho$,  are isomorphic in the context  $[~]$, with $\lambda xy.xy$ showing the isomorphism. But the same types are not isomorphic in the context $[~] \wedge \varphi$, because no $\eta$-expansion of the identity can map an atomic type into itself.

To formalise this notion,  \emph{paths of type contexts} are useful (Definition \ref{pd}). The path of a context describes which arrows need to be traversed in order to reach the hole, if it is possible, i.e. when there are no atoms on the way.   It is handy to have a notion of {\em agreement of a type with a path} (Definition \ref{atpe}(\ref{atpe3})), in order to assure that the types which are composed by intersection or union with the type context do not block the transformation. An intersection or  a union agrees with a path only if all types belonging to the intersection or to the union agree with that path.

In paths the symbol $\lp$  represents going down to the left of an arrow and the symbol  $\rp$ represents going down to the right of an arrow. For distribution rules it is enough to reach the hole, while for splitting rules one more arrow needs to be traversed. So two kinds of paths are useful. They are dubbed d-paths and s-paths, being used in distribution and splitting rules, respectively.
An s-path is a d-path terminated by the symbol $\ep$.

The agreement of a type with a set of d-paths (Definition \ref{atpe}(\ref{atpe4})) and the concatenation of d-paths (Definition \ref{atpe}(\ref{atpe5})) are useful for defining the erasure rules (Definition \ref{e}).

\begin{definition}\labelx{atpe}
\begin{enumerate}
\item\labelx{atpe1} A {\em d-path} $\q$  is a possibly empty string on the alphabet $\set{\lp,\rp}$.
\item\label{atpe2} An {\em s-path} $\q$  is a d-path followed by  $\ep$.
\item\labelx{atpe3} The {\em agreement} of a type $\tS$ with a d-path or an  s-path $\q$ (notation $\ag\tS\q$)  is the smallest relation between types and d-paths (s-paths) such that:\\
\centerline{$\begin{array}{lcl}
 \ag{\tA}{\epsilon}\text{ for all }\tA;&\qquad& \ag{\tB\to\tC}{\ep}\text{  for all }\tB,\tC;\\
 \ag\tT{\q}\text{  implies }\ag{\tT\to\tR}{\lp\q};&&  \ag\tR{\q}\text{  implies }\ag{\tT\to\tR}{\rp\q};\\
 \ag\tT{\q}\text{  and }\ag\tR{\q}\text{  imply }\ag{\tT\wedge\tR}{\q};&&
 \ag\tT{\q}\text{  and }\ag\tR{\q}\text{ imply }\ag{\tT\vee\tR}{\q}.
\end{array}$}
\item\labelx{atpe4} A  type $\tS$ {\em agrees} with a set of d-paths $\calP$ (notation $\ag\tS\calP$) if it agrees with all the d-paths in $\calP$, i.e.
$\ag\tS\q$ for all $\q\in\calP$.
\item\labelx{atpe5} If $\p $ and $\p' $  are d-paths, $\p \cdot \p'$  denotes their {\em concatenation}; if $\calP$ is a set of d-paths, $\p \cdot \calP$ denotes the set $\set{\p \cdot \q'\mid \q'\in\calP}\cup\set\p$.
\end{enumerate}
\end{definition}
\noindent
For example the type $\tS_{1}\to(\tS_{2}\to\tR_{1}\wedge\tR_{2})\wedge(\tS_{3}\vee \tS_{1}\to\tT_{1})\to\tT_{2}$ agrees with the d-path $\rp\lp\lp$ and with the s-path $\rp\lp\ep$, while
the type $\tS_{1}\to(\tS_{2}\to \tR_{1}\wedge\tR_{2})\wedge(\tS_{3}\vee \tS_{1}\to\tT_{1})\wedge\vf\to\tT_{2}$ agrees with the d-path $\rp\lp$ and with the s-path $\rp\ep$, but it does not agree with the d-path $\rp\lp\lp$ nor with the s-path $\rp\lp\ep$, since $\vf$ does not agree with $\lp$ nor with $\ep$.

\medskip

The d-paths and s-paths of contexts can be formalised using the agreement between types and paths.

\begin{definition}\labelx{pd}
The d-path and the s-path  of a type context $\C{~}$ (notations
$\pad{\C{~}}$ and $\pa{\C{~}}$, respectively) are  defined by
induction on $\C{~}$:\\[1mm]
\centerline{$\begin{array}{c}
 \pad{\C{~}}=\epsilon\text{ if }\C{~}=[~];\qquad
 \pa{\C{~}}=\ep\text{ if }\C{~}=[~];\\[1mm]
 *({\Cp{~}})=\p\text{ implies }*({\C{~}})=\lp\p\text{ if }\C{~}=\Cp{~}\to\tS \text{ and }
 *({\C{~}})=\rp\p\text{ if }\C{~}=\tS\to\Cp{~};\\[1mm]
\ag{\tS}{*({\Cp{~}})}\text{ implies } *({\C{~}})=*({\Cp{~}})\text{ if }\C{~} = \Cp{~}\wedge\tS\text{ or  }
\C{~} = \tS\wedge\Cp{~}\text{ or  }\\\phantom{\ag{\tS}{*({\Cp{~}})}\text{ implies } *({\C{~}})=*({\Cp{~}})}\C{~} = \Cp{~}\vee\tS \text{ or  }\C{~} = \tS\vee\Cp{~}.\\[1mm]
\end{array}$}
where $*$  holds for $d$ and $s$.
\end{definition}
\noindent
For example the d-path and the s-path of the context
$\tS_1\to[~]\wedge(\tS_2\to\tT_1\vee  \tT_{2})\to\tT_2$ are
$\rp\lp$ and $\rp\lp\ep$, respectively, while the d-path and the s-path of the context
$\tS_1\to([~]\wedge\tS_2\to\tT_1)\vee \vf\to\tT_2$ are undefined,
since $\nag\vf{\lp}$ and  $\nag\vf{\ep}$.

\medskip
In giving the  normalisation rules one can consider types in holes modulo idempotence, commutativity and associativity, when the d-paths of contexts are defined.
This is assured by the following lemma, that can be easily proved by induction on  d-paths.
\begin{lemma} \labelx{aci}
If $\tA\approx\tB$ holds the isomorphisms \mbox{(\rm{{\bfseries idem}})}, \mbox{(\rm{{\bfseries comm}})},    \mbox{(\rm{{\bfseries assoc}})}, and $ \pad{\C{~}}$ is defined, then $\C\tA\approx\C\tB$.
\end{lemma}

\medskip

Distribution and splitting rules can now be defined.
\begin{definition}[Distribution and Splitting]\labelx{ds}
\begin{enumerate}
\item
 The two {\em distribution rules} are:\\[3pt]
\centerline{$\C{(\tS\wedge\tT)\vee\tR}\red\C{(\tS\vee\tR)\wedge(\tT\vee\tR)}\qquad$ if $\pad{\C{~}} = \epsilon $ or $\pad{\C{~}} = \p \cdot \rp$ for some path $\p$;}
\centerline{$\C{(\tS\vee\tT)\wedge\tR}\red\C{(\tS\wedge\tR)\vee(\tT\wedge\tR)}\qquad$ if $\pad{\C{~}} = \p \cdot \lp $ for some path $\p.$\phantom{$\pad{\C{~}} = \epsilon $ or }}
\item
The two {\em splitting rules} are:\\[3pt]
\centerline{$\C{\tS\to\tT\wedge\tR}\red\C{(\tS\to\tT)\wedge(\tS\to\tR)}\qquad$ if $\pa{\C{~}}$ is defined;}
\centerline{$\C{\tS\vee\tT\to\tR}\red\C{(\tS\to\tR)\wedge(\tT\to\tR)}\qquad$ if $\pa{\C{~}}$ is defined.}
\end{enumerate}
\end{definition}

\medskip

The conditions for erasure rules use two preorders on types,  defined in Figure \ref{tpd} between basic intersections and  between basic unions (see page \pageref{basic}), respectively. This is enough since
the distribution and splitting rules (when applicable) give arrow types with basic intersections as left-hand-sides and basic unions as right-hand-sides.
The symbol $\md$ stands for either  $\mi$ or  $\muu$. It easy to verify that $\alpha\mi\beta$ if and only if $\alpha\muu\beta$,
so comparing two arrows or two atomic types one can write $\alpha\md\beta$.
%
%
For example $\tM\wedge\tN \mi \tM$ and $\ti \muu \ti\vee\tk$
imply $\tM\to\ti\md\tM\wedge\tN\to\ti\vee\tk$ and
$(\tM\wedge\tN\to\ti\vee\tk)\to\tx\md(\tM\to\ti)\to\tx$.

It is easy to show that $\mi$ and $\muu$ are preorders since transitivity holds.
The presence, at top level, of an atomic type on both sides of $\md$ forces atomic and arrow types to be only erased or added.
In relating types one can exploit also  idempotence. For instance two copies of $(\tM\to\ti)$ are needed in deriving $\tM\to\ti \mi (\tM\wedge\tN \to \ti)\wedge (\tM \to \ti\vee\tk)$ to show  $\tM\to\ti \md \tM\wedge\tN \to \ti$ and $\tM \to \ti \md \tM \to \ti\vee\tk$.
\begin{figure}[h]
\centerline{$\begin{array}{c}
 \tM  \mi  \tM
              \qquad  \ti \muu  \ti\qquad \varphi\wedge\tM \mi \varphi\qquad \varphi \muu \varphi\vee\ti
            \\[1mm]
            \varphi\wedge\tM\wedge\tl \mi \varphi\wedge\tM \qquad \varphi\vee\ti \muu \varphi\vee\ti\vee\tx \\[1mm]
\tN_i \mi  \tM_i, \;  \ti_i \muu  \tk_i \text{ for all $i\in I$ }
\Rightarrow
             \bigwedge_{i\in I}(\tM_i \to  \ti_i) [\wedge\tl] \mi  \bigwedge_{i\in I}(\tN_i \to  \tk_i) \\[1mm]
\tN_i \mi  \tM_i, \;  \ti_i \muu  \tk_i \text{ for all $i\in I$ }
\Rightarrow
             \bigvee_{i\in I}(\tM_i \to  \ti_i)  \muu  \bigvee_{i\in I}(\tN_i \to  \tk_i) [\vee\tx]\\[1mm]
\end{array}$}
 where the notation $[\wedge\tl]$ ($[\vee\tx]$) means that $\wedge\tl$ ($\vee\tx$) can either occur or not.
 \caption{Preorders on types.}\labelx{tpd}
\end{figure}
\begin{figure}[h]
\centerline{$\begin{array}{c}
\Sp(\tM \mi  \tM) = \Sp(\ti \muu  \ti)= \set{} \\[3mm]
 \Sp(\varphi\wedge\tM \mi  \varphi) = \Sp( \varphi \muu \varphi\vee\ti)=\Sp(\varphi\wedge\tM\wedge\tl \mi \varphi\wedge\tM) = \Sp( \varphi\vee\ti \muu \varphi\vee\ti\vee\tx)=\set{\epsilon}\\[3mm]
  \begin{array}{l}
 \begin{array}{l}
\Sp(           \bigwedge_{i\in I}(\tM_i \to  \ti_i)[\wedge \tl]  \mi  \bigwedge_{i\in I}(\tN_i \to  \tk_i))\\
\Sp(
             \bigvee_{i\in I}(\tM_i \to  \ti_i)   \muu  \bigvee_{i\in I}(\tN_i \to  \tk_i)[\vee \tx])
             \end{array}
         =  \begin{cases}
  \set{\epsilon}  \quad\quad   \text{if $\tl$ or $\tx$ is present and}\\
    \phantom{\set{\epsilon}  \quad\quad  }\text{$\Sp(\tN_i \mi  \tM_i)=\Sp(\ti_i \muu  \tk_i)=\set{}$ for all $i\in I$}, \\
   \bigcup_{i\in I} ( \lp\cdot \Sp(\tN_i \mi  \tM_i)\cup
    \rp\cdot \Sp(\ti_i \muu  \tk_i) ) \quad\text{otherwise}
\end{cases}\\
\quad\text{ if }\tN_i \mi  \tM_i, \;  \ti_i \muu  \tk_i \text{ for all $i\in
I$ }
\end{array}
   \end{array}$}
\caption{Set of d-paths of a preorder derivation.}\labelx{Spd}\end{figure}

These preorders are crucial for the definition of the erasure rules. In fact some types in an intersection can be erased only if the remaining types are smaller or equal to the erased ones. Dually
some types in a union can be erased only if the remaining types are bigger or equal to the erased ones. Another necessary condition for erasing types is that the \fhp s can reach the subtypes in which the types related by the preorder differ. In order to formalise this, one d-path is not enough, since there can be many subtypes in which the types differ, so sets of d-paths are needed.
Sets of d-paths are then associated with derivations of preorders
between types, so that one can check when a type can be erased in a
type context.
The set of d-paths of $\tA\md\tB$ (notation $\Sp(\tA\md\tB)$)
represents the set of paths that  make  accessible the points in which $\tA$ and $\tB$ differ.
For this reason, $\Sp(\tM\mi\tM)$ and $\Sp(\ti\muu\ti)$ are defined as the empty set and the sets:\\
 \centerline{$\Sp(\varphi\wedge\tM \mi  \varphi), \; \Sp( \varphi \muu \varphi\vee\ti),\; \Sp(\varphi\wedge\tM\wedge\tl \mi \varphi\wedge\tM), \;$ and $\; \Sp( \varphi\vee\ti \muu \varphi\vee\ti\vee\tx)$} contain  only $\epsilon$;
in the other cases this set must be built from the sets of paths associated with the subtypes using $\lp$ and $\rp$. This definition is given in Figure \ref{Spd}. Notice that the condition $\Sp(\tA \md  \tB) = \set{}$ implies $\tA = \tB$.

\noindent
For example
$\Sp(\tM\to\ti\md\tM\wedge\tN\to\ti\vee\tk)=\set{\lp,\rp}$ and
$\Sp((\tM\wedge\tN\to\ti\vee\tk)\to\tx\md(\tM\to\ti)\to\tx)=\set{\lp\lp,\lp\rp}$.

\medskip

Finally one can define erasure rules.

\begin{definition}[Erasure]\labelx{e}
The three {\em erasure rules} are:\\[3pt]
\centerline{$\begin{array}{ll}
\bigwedge_{i\in I}\ti_i\red \bigwedge_{j\in J}\ti_j & \text{ if $J\subset I$ and $\forall i\in I\; \exists j_i\in J.\; \ti_{j_i}\muu \ti_i$  and $\forall j\in J.\; \ag{\ti_j} {\cal{P}}$,}\\
&\text{where $\calP = \bigcup_{i \in I}\Sp(\ti_{j_i}\muu\ti_i)$};
\\[3pt]
\C{\bigwedge_{i\in I}\tO_i}\red\C{\bigwedge_{j\in J}\tO_j} & \text{ if $J\subset I$ and $\forall i\in I\; \exists j_i\in J.\; \tO_{j_i}\md \tO_i$ and $\forall j\in J.~ \ag{C[\tO_j]} {\calP}$},\\
 &\text{where $\calP = d(C[])\,\cdot\,\bigcup_{i \in I}\Sp(\tO_{j_i}\md\tO_i)$};
\\[3pt]
\C{\bigvee_{i\in I}\tO_i}\red\C{\bigvee_{j\in J}\tO_j} & \text{ if $J\subset I$ and $\forall i\in I\; \exists j_i\in J.\; \tO_i\md \tO_{j_i}$ and $\forall j\in J.~ \ag{C[\tO_j]} {\calP}$},\\
 &\text{where $\calP = d(C[])\,\cdot\,\bigcup_{i \in I} \Sp(\tO_i\md\tO_{j_i})$}.
 \end{array}$}
\end{definition}
\noindent
In the first erasure rule the absence of the context indicates that it can be applied only at top level, i.e. in the empty context.

By applying the erasure rules, it is essential to allow to remove more than one type in a single step. For example $(\tM\to\varphi\to\ti)\wedge(\tM\to(\varphi\wedge\tN\to\ti)\vee\psi_1)\wedge (\tM\to(\varphi\wedge\tN\to\ti)\vee\psi_2)\red\tM\to\varphi\to\ti$,
but this type does not reduce to $(\tM\to\varphi\to\ti)\wedge(\tM\to(\varphi\wedge\tN\to\ti)\vee\psi_i)$ for $i=1$ or $i=2$. The problem is that  $\tM\to(\varphi\wedge\tN\to\ti)\vee\psi_1$ does not agree with \mbox{$\Sp(\tM\to\varphi\to\ti\md\tM\to(\varphi\wedge\tN\to\ti)\vee\psi_2)=\set{\rp\lp}$} and dually exchanging $\psi_1$ with $\psi_2$.

Normalisation can create redexes, for example the first distribution rule applied to $\tA\to(\tB\wedge\tC)\vee\tD$ gives $\tA\to(\tB\vee\tD)\wedge(\tC\vee\tD)$, which can be reduced to $(\tA\to\tB\vee\tD)\wedge(\tA \to \tC\vee\tD)$ by the first splitting rule. 
The second splitting rule applied to $(\tA\vee\varphi\to\varphi)\wedge(\varphi\wedge\psi\to\varphi)$ gives  $(\tA\to\varphi)\wedge(\varphi\to\varphi)\wedge(\varphi\wedge\psi\to\varphi)$, which can be reduced to $(\tA\to\varphi)\wedge(\varphi\to\varphi)$ by the first or second erasure rule. A more interesting example is $(\varphi\wedge(\psi\to\psi)\to\psi)\wedge((\psi\to\psi)\to\psi)\wedge(((\tA\vee\tB)\wedge\tC\to\tC)\to\tC)$: this type can only be reduced to $((\psi\to\psi)\to\psi)\wedge(((\tA\vee\tB)\wedge\tC\to\tC)\to\tC)$ by the first or second erasure rule and then the second distribution rule becomes applicable.

\subsection{Soundness, confluence and termination of type normalisation}\labelx{AB}\spazios

The soundness of the normalisation rules, i.e. that $\tA\red\tB$ implies $\tA\approx\tB$, uses $\eta$-expansions of the identity, called {\em finite hereditarily identities  (\fhi s)}. More precisely
for each rule $\tA\red\tB$ two \fhi s $\id, \id'$ such that $\vdash\id\dup\tA\to\tB$ and $\vdash\id'\dup\tB\to\tA$ are built.
\fhi s can be associated with d-paths, s-paths and sets of d-paths.
\begin{definition}\labelx{fhip}
\begin{enumerate}
\item\labelx{fhip1} The \fhi\ induced by the s-path $\p$ (notation $\id_\p$) is defined by induction on $\p$:\\
\centerline{$\id_{\ep}=\lambda xy.xy \qquad
\id_{\lp\p}\tob\lambda xy.x(\id_{\p} y)\qquad
\id_{\rp\p}\tob\lambda xy.\id_{\p}(xy)$}
\item\labelx{fhip2} The \fhi\  induced by the set of d-paths $\calP$ (notation $\id_\calP$) is defined by induction on the d-paths in $\calP$:
\centerline{$\id_{\set{\;}}=\id_{\set{\epsilon}}=\lambda x.x \qquad
\id_{\calP}\tob\lambda xy.\id_{\calR(\calP)}(x(\id_{\calL(\calP)}y))\text{ if }\calP\not=\set{\;},\set{\epsilon}$}
where $\calL(\calP)=\set{\q\mid \lp\q\in \calP}$ and $\calR(\calP)=\set{\q\mid \rp\q\in \calP}$.
\item\labelx{fhip3} The \fhi\  induced by the d-path $\q$ (notation $\id_\q$) is $\id_\q=\id_{\set\q}$.
\end{enumerate}
\end{definition}
\noindent
For example
$\id_{\rp\lp\ep}\tob\lambda x_1y_1.\id_{\lp\ep}(x_1y_1)$
$ \tob\lambda x_1y_1.(\lambda x_2y_2.x_2(\id_{\ep}y_2))(x_1y_1)$
$\tob$\\
$\lambda x_1y_1.(\lambda x_2y_2.x_2((\lambda x_3y_3.x_3y_3)y_2))(x_1y_1)$,
so $\id_{\rp\lp\ep}$ $=$ $\lambda x_1y_1y_2.x_1y_1(\lambda y_3.y_2y_3)$.

\medskip

The following lemma  shows that the \fhi\  associated with a d-path,
an s-path or a set of d-paths maps to itself each type that
agrees with it.

\begin{lemma}\labelx{pl}
\begin{enumerate}
\item\labelx{pl1} Let $\q$ be a d-path or an s-path, then $\ag\tS\q$ implies $\der{}{\id_\q}{\tS\to\tS}$.
\item\labelx{pl2} Let $\calP$ be a set of d-paths, then $\ag\tS\calP$ implies $\der{}{\id_\calP}{\tS\to\tS}$.
\end{enumerate}
\end{lemma}\spaziop
\begin{proof}
Only Point
(\ref{pl2}) is proved, being the proof of Point (\ref{pl1}) similar and simpler. The proof is by induction on $\tS$ and $\calP$. If
$\tS=\tT\to\tR$, then by definition $\ag\tT{\calL(\calP)}$ and
$\ag\tR{\calR(\calP)}$. By induction
$\der{}{\id_{\calL(\calP)}}{\tT\to\tT}$ and
$\der{}{\id_{\calR(\calP)}}{\tR\to\tR}$, which imply
$\der{}{\lambda
xy.\id_{\calR(\calP)}(x(\id_{\calL(\calP)}y))}{\tS\to\tS}$. If
$\tS=\tT\vee\tR$ or $\tS=\tT\wedge\tR$, then by definition
$\ag\tT\calP$ and $\ag\tR\calP$. These cases easily follow by
induction using Corollary~\ref{cor}(\ref{cor1}).
\end{proof}

To prove the soundness of erasure one needs to show that the \fhi\
associated with a set of d-paths ``respects" the preorder relation,
in the sense that, if the set of d-paths of the derivation $\tA \md
\tB$ is contained in a set $\calP$ and either $\tA$ or $\tB$ agrees with $\calP$,
then the \fhi\  $\id_\calP$ maps $\tA$ to $\tB$.

\begin{lemma}\labelx{keye}
If $\Sp(\tS\md\tT)\subseteq\calP$ and either $\ag{\tS}\calP$ or $\ag{\tT}\calP$, then $\der{}{\id_\calP}{\tS\to\tT}$.
\end{lemma}\spaziop
\begin{proof}
By induction on the proof of $\tS\md\tT$. The cases $\tM \mi  \tM$ and $\ti \muu  \ti$ follow immediately by Lemma \ref{pl}.
  In cases $\varphi\wedge\tM\mi\varphi, \varphi\muu\varphi\vee\ti, \varphi\wedge\tM\wedge\tl \mi \varphi\wedge\tM$ and $\varphi\vee\ti \muu \varphi\vee\ti\vee\tx$ one has $\Sp(\tS\md\tT) = \set{\epsilon}$. Since $\ag{\tS}\calP$ or $\ag{\tT}\calP$ implies $\ag{\varphi}{\calP}$, ~
   $\calP = \set{\epsilon}$ and by Definition \ref{fhip}(\ref{fhip2}), one gets ${\id_\calP} = \lambda x. x$. \\
Consider the case: $\tN_i \mi  \tM_i, \;  \ti_i \muu  \tk_i \text{ for all $i\in I$ } \; \Rightarrow \; \bigwedge_{i\in I}(\tM_i \to  \ti_i)[\wedge\tl ] \mi  \bigwedge_{i\in I}(\tN_i \to  \tk_i)$. By definition:
\begin{itemize}
\item $\Sp(\bigwedge_{i\in I}(\tM_i \to  \ti_i)[\wedge\tl ] \mi  \bigwedge_{i\in I}(\tN_i \to  \tk_i))\subseteq\calP$ implies $\Sp(\tN_i \mi  \tM_i)\subseteq\calL(\calP)$ and $\Sp(\ti_i \muu  \tk_i )\subseteq\calR(\calP)$ for all $i\in I$;
 \item either $\ag{\bigwedge_{i\in I}(\tM_i \to  \ti_i)[\wedge\tl ] }\calP$ or $\ag{\bigwedge_{i\in I}(\tN_i \to  \tk_i)}\calP$ implies either $\ag{\tN_i }{\calL(\calP)}$ or $\ag{\tM_i}{\calL(\calP)}$ and either $\ag{\ti_i }{\calR(\calP)}$ or $\ag{\tk_i}{\calR(\calP)}$ for all $i\in I$.
\end{itemize}
This gives by induction $\der{}{\id_{\calL(\calP)}}{\tN_i \to  \tM_i}$ and $\der{}{\id_{\calR(\calP)}}{\ti_i \to  \tk_i}$ for all $i\in I$. By definition  \\ $\id_{\calP}\tob\lambda xy.\id_{\calR(\calP)}(x(\id_{\calL(\calP)}y))$. It can be easily shown that:\\
 \centerline{$\der{}{\lambda xy.\id_{\calR(\calP)}(x(\id_{\calL(\calP)}y))}{(\tM_i \to  \ti_i)\to\tN_i \to  \tk_i}$ for all $i\in I$.}\\
The Subject Reduction (Theorem \ref{srtheorem}) implies $\der{}{\id_\calP}{(\tM_i \to  \ti_i)\to\tN_i \to  \tk_i}$ for all $i\in I$, and so by
Corollary~\ref{cor}(\ref{cor1}) and (\ref{cor4})\\
\centerline{$\der{}{\id_\calP}{\bigwedge_{i\in I}(\tM_i \to  \ti_i)[\wedge\tl]  \to  \bigwedge_{i\in I}(\tN_i \to  \tk_i)}$.}
For the case: $\tN_i \mi  \tM_i, \;  \ti_i \muu  \tk_i \text{ for all $i\in I$ } \; \Rightarrow \;
             \bigvee_{i\in I}(\tM_i \to  \ti_i)  \muu  \bigvee_{i\in I}(\tN_i \to  \tk_i)[\vee\tx]$ a similar argument gives
 $\der{}{\id_\calP}{\bigvee_{i\in I}(\tM_i \to  \ti_i)  \to  \bigvee_{i\in I}(\tN_i \to  \tk_i)[\vee\tx]}$.
\end{proof}

The soundness of the normalisation rules can now be proved.

\begin{theorem}\labelx{spl}
\begin{enumerate}
\item\labelx{spl1} If $\pad{\C{~}}$ is defined, then  for arbitrary $\tS$, $\tT$, and $\tR$, the \fhi\  $\id_{\pad{\C{~}}}$ proves the isomorphisms:
 $\C{(\tS\vee\tT)\wedge\tR}\isot\C{(\tS\wedge\tR)\vee(\tT\wedge\tR)}$ \;\; and\;\;
 $\C{(\tS\wedge\tT)\vee\tR}\isot\C{(\tS\vee\tR)\wedge(\tT\vee\tR)}$.
\item\labelx{spl2} If $\pa{\C{~}}$ is defined, then for arbitrary $\tS$, $\tT$, and $\tR$, the \fhi\  $\id_{\pa{\C{~}}}$ proves the isomorphisms:\\
  $\C{\tS\to\tT\wedge\tR}\isot\C{(\tS\to\tT)\wedge(\tS\to\tR)}$ \;\; and\;\;
  $\C{\tS\vee\tT\to\tR}\isot\C{(\tS\to\tR)\wedge(\tT\to\tR)}$.
\item\labelx{spl3}Let $\bigwedge_{i\in I}\ti_i\red \bigwedge_{j\in J}\ti_j$, i.e.
    $J\subset I$ and $\forall i\in I\; \exists j_i\in J.\; \ti_{j_i}\muu \ti_i$  and   $\forall j\in J.~ \ag{\ti_j} {\cal{P}}$,\\  where $\calP={\bigcup_{i\in I}\Sp(\ti_{j_i}\muu \ti_i)}$.
      Then $\id_\calP$ proves $\bigwedge_{j\in J}\ti_j\isot\bigwedge_{i\in I}\ti_i$.
%
\item\labelx{spl5}   Let $\C{\bigwedge_{i\in I}\tO_i}\red\C{\bigwedge_{j\in J}\tO_j}$, i.e. $J\subset I$ and $\forall i\in I\; \exists j_i\in J.\; \tO_{j_i}\md \tO_i$  and $\forall j\in J.~ \ag{C[\tO_j]} {\calP}$,\\
 where $\calP = d(C[])\;\cdot\;\bigcup_{i \in I}\Sp(\tO_{j_i}\md\tO_i)$. Then
 $\id_\calP$ proves
 $\C{\bigwedge_{j\in J}\tO_j}\isot\C{\bigwedge_{i\in I}\tO_i}$.
 %
\item\labelx{spl6}  Let $\C{\bigvee_{i\in I}\tO_i}\red\C{\bigvee_{j\in J}\tO_j}$ ~ i.e. $J\subset I$ and $\forall i\in I.\; \exists j_i\in J.\; \tO_i\md \tO_{j_i}$  and $\forall j\in J.\; \ag{C[\tO_j]} {\calP}$,\\
  where $\calP = d(C[])\;\cdot\;\bigcup_{i \in I} \Sp(\tO_i\md\tO_{j_i})$.
  Then
 $\id_\calP$ proves
 $\C{\bigvee_{i\in I}\tO_i}\isot\C{\bigvee_{j\in J}\tO_j}$.

\end{enumerate}
\end{theorem}\spaziop
\begin{proof}
(\ref{spl1}). By induction on $\C{~}$. If $\C{~}=[~]$ by Definition \ref{pd} $\pad{[~]}=\epsilon$ and $\id_{\epsilon}=\lambda x.x$.

If $\C{~}=\Cp{~}\to\tD$, then by induction\\
\centerline{$\begin{array}{l}
\der{}{\id_{\pad{\Cp{~}}}}{\Cp{(\tS\vee\tT)\wedge\tR}\to\Cp{(\tS\wedge\tR)\vee(\tT\wedge\tR)}}\\
\der{}{\id_{\pad{\Cp{~}}}}{\Cp{(\tS\wedge\tR)\vee(\tT\wedge\tR)}\to\Cp{(\tS\vee\tT)\wedge\tR}}.
\end{array}$}
Since by definition $\id_{\pad{\C{~}}}\tob\lambda xy.x(\id_{\pad{\Cp{~}}}y)$ the result follows.

If $\C{~}=\tD\to\Cp{~}$ the proof is similar to that one of previous case.

If $\C{~}=\Cp{~}\wedge\tD$, then by induction\\
\centerline{$\begin{array}{l}
\der{}{\id_{\pad{\Cp{~}}}}{\Cp{(\tS\vee\tT)\wedge\tR}\to\Cp{(\tS\wedge\tR)\vee(\tT\wedge\tR)}}\\
\der{}{\id_{\pad{\Cp{~}}}}{\Cp{(\tS\wedge\tR)\vee(\tT\wedge\tR)}\to\Cp{(\tS\vee\tT)\wedge\tR}}.
\end{array}$}
Moreover $\pad{\C{~}}=\pad{\Cp{~}}$ and $\ag{\tD}{\pad{\Cp{~}}}$, so from Lemma~\ref{pl}(\ref{pl1})   $\der{}{\id_{\pad{\Cp{~}}}}{\tD\to\tD}$
and by Corollary~\ref{cor}(\ref{cor1})  the proof is done.\\
If $\C{~}=\Cp{~}\vee\tD$, the proof is similar.

(\ref{spl2}). Similar to the proof of Point (\ref{spl1}).
The only difference is case $\C{~}=[~]$, in which by Definition \ref{pd}
$\pa{[~]}=\ep$ and $\id_{\ep}=\lambda xy.xy$.

(\ref{spl3}). Lemma \ref{keye} implies $\der{}{\id_\calP}{\ti_{j_i} \to \ti_i}$, since $\ag{\ti_{j_i}}{\calP}$ for all $i\in I$, and $\calP={\bigcup_{i\in I}\Sp(\ti_{j_i}\muu \ti_i)}$. Lemma \ref{pl}(\ref{pl2}) gives  $\der{}{\id_\calP}{\ti_{j} \to \ti_j}$ for all $j\in J$, since $\ag{\ti_j}{\calP}$ for all $j\in J$.
So, by Corollary \ref{cor}(\ref{cor1}), $\id_\calP$ has both the types $\bigwedge_{i\in I}\ti_{j_i}\to\bigwedge_{i\in I}\ti_i$ and $\bigwedge_{j\in J}\ti_j\to\bigwedge_{j\in J}\ti_j$. Finally, Corollary \ref{cor}(\ref{cor4}) implies  \mbox{$\der{}{\id_\calP} {\bigwedge_{j\in J}\ti_j}\to\bigwedge_{i\in I}\ti_i$.}

(\ref{spl5}). By induction on $\C{\;}$. If $\C{\;}=[~]$, the proof is immediate from Point (\ref{spl3}).\\
Let $\calP ' = d(C'[])\,\cdot\,\bigcup_{i \in I}\Sp(\tO_{j_i}\md\tO_i)$.

 If $\C{~}=\Cp{~}\to\tS$, then $\pad{\C{~}}= \lp \cdot \pad{\Cp{~}}$. By induction\\
\centerline{$
\der{}{\id_{\calP'}}
{\Cp{\bigwedge_{j\in J}\tO_j} \to \Cp{\bigwedge_{i \in I} \tO_i}}$ \;and\; $\der{}{\id_{\calP '}}
{\Cp{\bigwedge_{i \in I} \tO_i}\to\Cp{\bigwedge_{j \in J}  \tO_j}}$}.
Since by definition $\id_{\calP} \tob\lambda xy.x(\id_{\calP '} y)$, the result follows.

If $\C{~}=\tS\to\Cp{~}$ the proof is similar to that one of previous case.

If $\C{~}=\Cp{~}\wedge\tS$, then by induction\\
\centerline{$
\der{}{\id_{\calP '}}
{\Cp{\bigwedge_{j\in J}\tO_j} \to \Cp{\bigwedge_{i \in I} \tO_i}}$ \;and\; $\der{}{\id_{\calP '}}{\Cp{\bigwedge_{i \in I} \tO_i}\to\Cp{\bigwedge_{j \in J}  \tO_j}}$}.
In this case $\calP=\calP'$ and $\ag{\tS}{\calP}$. Lemma~\ref{pl}(\ref{pl2}) gives $\der{}{\id_{\calP}}{\tS\to\tS}$ and Corollary~\ref{cor}(\ref{cor1}) concludes  the proof.


If $\C{~}=\Cp{~}\vee\tS$, the proof is similar.

(\ref{spl6}). Similar to the proof of Point (\ref{spl5}).
\end{proof}

This subsection ends with
the proof of the existence and unicity of normal forms, i.e. that the normalisation rules are terminating and confluent.
\begin{theorem}[Normal Forms]\labelx{confl}
The rewriting system of Definitions \ref{ds}  and \ref{e} is terminating and confluent.
\end{theorem}\spaziop
\begin{proof}The {\em termination} follows from an easy adaptation of the recursive path ordering method \cite{D82}. The partial order on operators is defined by: $\to~\succ~\vee~\succ~\wedge$ for  holes at top level or in the right-hand-sites of arrow types and  $\to~\succ~\wedge~\succ~\vee$ for holes in the left-hand-sites of arrow types. Notice that the induced recursive path ordering $\succ^*$ has the subterm property. This solves the case of erasure rules.
For the first distributive rule, since $\vee~\succ~\wedge$ for holes at top level or in the right-hand-sites of arrow types, it is enough to observe that $(\tS\wedge\tT)\vee\tR~\succ^*~\tS\vee\tR$ and        $(\tS\wedge\tT)\vee\tR~\succ^*~\tT\vee\tR$. For the first splitting rule, since $\to~\succ~\wedge$, it is enough to observe that $\tS\to\tT\wedge\tR~\succ^*~\tS\to\tT$ and        $\tS\to\tT\wedge\tR~\succ^*~\tS\to\tR$.
The proof for the remaining rules are similar.

 For {\em confluence}, following the Knuth-Bendix algorithm \cite{KB70} it is sufficient to  prove  the convergence of the critical
 pairs, that are generated  (modulo commutativity and associativity of union and intersection) by:\\
  \centerline{$(\bigwedge_{i\in I}\tO_i)
\vee \tA $,\qquad $ \tA \to
(\bigwedge_{i\in I}\tO_i) \vee \tB$,\qquad$(\bigvee_{i\in I}\tO_i) \wedge \tA \to \tB$,}\\  \centerline{$ \tA \to
(\bigvee_{i\in I}\tO_i) \wedge \tB$,\qquad$\tA \vee (\bigwedge_{i\in I}\tO_i)\to \tB$,\;\;$ \tA \to
\bigwedge_{i\in I}\tO_i \wedge \tB$, \qquad \;$\tA \vee \bigvee_{i\in I}\tO_i\to \tB$,}\\
\centerline{$(\tA\wedge\tB)\vee\tC\vee(\tD\wedge\tE)$,\qquad$(\tA\vee\tB)\wedge\tC\wedge(\tD\vee\tE)\to\tF$,}\\
\centerline{$\tA\vee\tB\vee\tC\to\tD,$\qquad$\tA\to\tB\wedge\tC\wedge\tD$,}\\
\centerline{$\bigwedge_{i\in I}\ti_i,$\qquad$\bigwedge_{i\in I}\tO_i$,\qquad$\bigvee_{i\in I}\tO_i$,}\\
when $\bigwedge_{j\in J}\tO_j \mi \bigwedge_{i\in I}\tO_i,\;\;\bigvee_{i\in I}\tO_i\muu \bigvee_{j\in J}\tO_j$, $J\subset I$ and all the  conditions required by the normalisation rules are satisfied.
The types of the last line generate critical pairs when, for some $L\subset I, \;L\neq J,$ one has, for the first type $\forall i\in I\; \exists j_i\in J\; l_i\in L$ such that $\ti_{j_i} \muu \ti_i$ and $\ti_{l_i} \muu \ti_i$ and, for the second and the third types, in addition to the above conditions, $\bigwedge_{l\in L}\tO_j \mi \bigwedge_{i\in I}\tO_i$ and $\bigvee_{l\in L}\tO_i\muu \bigvee_{j\in J}\tO_j$, respectively.
%

The types in the first line can be reduced by distribution and erasure rules, the types in the second line can be reduced by splitting and erasure rules, the types in the third line can be reduced by distribution rules, the types in the forth line can be reduced by splitting rules, the types in the last line can be reduced by erasure rules. Notice that distribution and splitting rules do not generate critical pairs, since they require respectively unions and intersections for holes in the right-hand-sites of arrow types and intersections and unions for holes in the left-hand-sites of arrow types.

The proof is given only for a top level occurrence of the type $(\bigwedge_{i\in I}\tO_i) \vee \tA $, the proof for the other cases being similar.
 In this case $(\bigwedge_{i\in I}\tO_i) \vee \tA \red (\bigwedge_{j\in J}\tO_j) \vee \tA$ by erasure since $J\subset I$ and $\forall i\in I\; \exists j_i\in J$ such that $\tO_{j_i}\md\tO_i$ and
 $\ag{(\bigwedge_{i\in I}\tO_i) \vee \tA}{\bigcup_{i\in I}\Sp(\tO_{j_i}\md\tO_i)}$. Moreover\\
 \centerline{$(\bigwedge_{i\in I}\tO_i) \vee \tA \red (\bigwedge_{i\in I_1}\tO_i \vee \tA) \wedge
(\bigwedge_{i\in I_2}\tO_i \vee \tA)$} with $I_1\cup I_2=I$ and $I_1\cap I_2=\emptyset$ by distribution. Let $\bigwedge_{l\in L} \ti_l$ be the conjunctive normal form of $\tA$, then $\tA\red^*\bigwedge_{l\in L} \ti_l$. This implies
$(\bigwedge_{j\in J}\tO_i) \vee \tA\red^*\bigwedge_{l\in L}\bigwedge_{j\in J}(\tO_i\vee \ti_l)$ and\\
\centerline{$(\bigwedge_{i\in I_1}\tO_i \vee \tA) \wedge(\bigwedge_{i\in I_2}\tO_i \vee \tA)\red^*\bigwedge_{l\in L}(\bigwedge_{i\in I_1}\tO_i\vee \ti_l)\wedge(\bigwedge_{i\in I_2}\tO_i\vee \ti_l)=\bigwedge_{l\in L}\bigwedge_{i\in I}(\tO_i\vee \ti_l)$} by the first distribution rule.
 Since $\tO_{j_i}\md\tO_i$ implies $\tO_{j_i} \vee \ti_l\muu \tO_i\vee \ti_l$, and $\ag{(\bigwedge_{j\in J}\tO_j) \vee \tA}{\bigcup_{i\in I}\Sp(\tO_{j_i}\md\tO_i)}$ implies $\ag{\bigwedge_{l\in L}\bigwedge_{j\in J}(\tO_j \vee \ti_l)}{\bigcup_{l\in L}\bigcup_{i\in I}\Sp(\tO_{j_i} \vee \ti_l\muu \tO_i\vee \ti_l)}$, the first erasure rule gives\\
 \centerline{$\bigwedge_{l\in L}\bigwedge_{i\in I}(\tO_i\vee \ti_l)\red\bigwedge_{l\in L}\bigwedge_{j\in J}(\tO_j\vee \ti_l)$.}\vspace{-15pt}  \end{proof}

 \subsection{Properties of normal types}\labelx{AD}\spazios
It is interesting to show that normal types do not contain ``superfluous'' subtypes, in particular that:
\begin{enumerate}
\item\labelx{s1} if $\ti\wedge\tk$ is a normal type, then there is no $\id$ such that $\vdash\id\dup\ti\to\tk$;
\item\labelx{s2} if $\tO\vee\tP$ is a normal type, then there is no $\id$ such that $\vdash\id\dup\tO\to\tP$;
\end{enumerate}
Theorem \ref{ntT1} shows Points (\ref{s1}) and (\ref{s2}). 


The more interesting result is Theorem \ref{discon}, which assures that isomorphic normal types have the same number of intersections and unions and that the atomic and arrow types are pairwise isomorphic.

\medskip

It is easy to verify that
each \fhi\  $\id$ different from the identity is such that  $\id\tob\lambda xy. \id_1(x(\id_2y))$ for unique $\id_1,\id_2$. A key result is a relation between the arrow types that can be derived for $\id, \id_1,\id_2$. Lemma \ref{first}(\ref{first3}) gives this relation, by exploiting the constraints on typings  of variables and \fhi s, shown in the first three points of the same lemma.

\medskip

Building on Definition \ref{fhip}(\ref{fhip2}) a non-empty set of d-paths is associated with each \fhi\ (Definition \ref{Ipd}). This association is based on the natural correspondence between lambda abstractions and arrow types.

\medskip

A last Lemma (Lemma \ref{nt1}) relates basic intersections and unions in normal form  (when they can be mapped by \fhi s) to preorders  and to sets of d-paths.

\begin{lemma}\labelx{first}
\begin{enumerate}
\item \labelx{first1} If $ \B,x\dup\tA\to\tB, y\dup\tC \vdash x(My)\dup\tD$, then $\B\vdash \lambda y. M y\dup\tC\to\tA$.
\item \labelx{first4}  Let $M$ be either a \fhi\ or a free variable. Then $\B,x\dup\tA\to\tB, y\dup\tC \vdash M(xy)\dup\tD$ implies $\B\vdash \lambda z. M z\dup\tB\to\tD$ and $z\dup\tC\vdash z\dup\tA$.
\item \labelx{first2}  Let $M_1,M_2$ be either \fhi s or free variables and $FV(M_i)$  be the set of variables in $\B_i$ for $i=1,2$. Then $\B_1, \B_2,x\dup\tA\to\tB, y\dup\tC \vdash M_1(x(M_2y))\dup\tD$ implies $\B_1\vdash \lambda z. M_1 z\dup\tB\to\tD$ and $\B_2\vdash\lambda z. M_2 z\dup\tC\to\tA$.
\item \labelx{first3} If $\vdash\id\dup(\tM\to\ti)\to\tN\to\tk$ and $\id\tob\lambda xy. \id_1(x(\id_2y))$, then $\vdash\id_1\dup\ti\to\tk$ and $\vdash\id_2\dup\tN\to\tM$.
\end{enumerate}
\end{lemma}\spaziop
\begin{proof}
(\ref{first1}). Lemma \ref{simpleM2}(\ref{simple1}) implies $\B,y\dup\tC \vdash My\dup\tA$, and rule $(\to I)$ derives $\B \vdash \lambda y. My\dup\tC\to\tA$.

(\ref{first4}). The proof is similar and simpler than that of (\ref{first2}).

(\ref{first2}). A stronger statement, i.e.\\[1mm] \centerline{\em $x\dup\tA\to\tB\vdash x\dup\tG$ and $\B_1, \B_2,x\dup\tG, y\dup\tC \vdash M_1(x(M_2y))\dup\tD$ imply $\B_1\vdash \lambda z. M_1 z\dup\tB\to\tD$ and $\B_2\vdash\lambda z. M_2 z\dup\tC\to\tA$,}\\[1mm] is proved by induction on the derivation of $\B_1, \B_2,x\dup\tG, y\dup\tC \vdash M_1(x(M_2y))\dup\tD$.

Let the last applied rule be $(\to E)$:
\[(\to E)\quad\frac{\B_1\vdash M_1\dup\tE\to\tD~~~~ \B_2,x\dup\tG, y\dup\tC \vdash x(M_2y)\dup\tE}{\B_1, \B_2,x\dup\tG, y\dup\tC \vdash M_1(x(M_2y))\dup\tD}\]
The second premise and $x\dup\tA\to\tB\vdash x\dup\tG$ imply $\B_2,x\dup\tA\to\tB, y\dup\tC \vdash x(M_2y)\dup\tE$ by rule $(L)$. Point (\ref{first1}) gives $\B_2 \vdash \lambda y. M_2y\dup\tC\to\tA$   and Lemma \ref{simpleM2}(\ref{simple2}) gives $z\dup\tB\vdash z\dup\tE$. The application of $(\to E)$ to the first premise and to $z\dup\tB\vdash z\dup\tE$ derives $ \B_1,z\dup\tB\vdash M_1z\dup\tD$, and then $\B_1\vdash \lambda z. M_1 z\dup\tB\to\tD$ by using $(\to I)$.

If the last applied rule is $(\wedge I)$, $(\wedge E)$, or $(\vee I)$ the proof easily follows by induction.


%
%
For rule $(\vee E)$ there are seven cases, which differ for the subjects of the premises. I.e. if $t$ is the replaced variable the subjects of the first two premises can be:
$t(x(M_2y))$, $M_1(t(M_2y))$, $M_1(x(ty))$, $M_1(x(M_2t))$, $M_1t$, $M_1(xt)$ and $t$.
The proof is given for  all the cases but the last one, which easily follows by induction. Notice that the proof of
 the sixth case needs Point (\ref{first4}).

In the first case:
\[(\vee E)\quad\frac{\begin{array}{c}\B_2,x\dup\tG, y\dup\tC, t\dup\tE_1\wedge\tF \vdash t(x(M_2y))\dup\tD~~~~ \B_2,x\dup\tG, y\dup\tC, t\dup\tE_2\wedge\tF \vdash t(x(M_2y))\dup\tD\\\B_1\vdash M_1\dup(\tE_1\vee\tE_2)\wedge\tF\end{array}}{\B_1, \B_2,x\dup\tG, y\dup\tC \vdash M_1(x(M_2y))\dup\tD}\]
By induction $t\dup\tE_1\wedge\tF\vdash \lambda z. t z\dup\tB\to\tD$ and $t\dup\tE_2\wedge\tF\vdash \lambda z. t z\dup\tB\to\tD$ and $\B_2 \vdash \lambda y. M_2y\dup\tC\to\tA$. By rule $(\vee L)$ $t\dup(\tE_1\vee\tE_2)\wedge\tF\vdash \lambda z. t z\dup\tB\to\tD$, so the application of rule $(C)$ to the third premise derives $\B_1\vdash \lambda z. M_1 z\dup\tB\to\tD$.

In the second case:
\[(\vee E)\quad\frac{\begin{array}{c}\B_1,\B_2,y\dup\tC, t\dup\tE_1\wedge\tF \vdash M_1(t(M_2y))\dup\tD~~~~\B_1,\B_2,y\dup\tC, t\dup\tE_2\wedge\tF \vdash M_1(t(M_2y))\dup\tD\\
x\dup\tG\vdash x\dup(\tE_1\vee\tE_2)\wedge\tF\end{array}}{\B_1, \B_2,x\dup\tG, y\dup\tC \vdash M_1(x(M_2y))\dup\tD}\]
Rule $(L)$ applied to $x\dup\tA\to\tB\vdash x\dup\tG$ and to the third premise derives $x\dup\tA\to\tB\vdash x\dup(\tE_1\vee\tE_2)\wedge\tF$. Corollary \ref{simpleM1}(\ref{ss1}) gives either $x\dup\tA\to\tB\vdash x\dup\tE_1\wedge\tF$ or $x\dup\tA\to\tB\vdash x\dup\tE_2\wedge\tF$. This implies  either $t\dup\tA\to\tB\vdash t\dup\tE_1\wedge\tF$ or $t\dup\tA\to\tB\vdash t\dup\tE_2\wedge\tF$. By induction on the first premise in the first case and on the second premise in the second case $\B_1\vdash \lambda z. M_1 z\dup\tB\to\tD$ and $\B_2\vdash\lambda z. M_2 z\dup\tC\to\tA$.

In the third case:
\[(\vee E)\quad\frac{\begin{array}{c}\B_1,x\dup\tG, y\dup\tC, t\dup\tE_1\wedge\tF \vdash M_1(x(ty))\dup\tD~~~~\B_1,x\dup\tG, y\dup\tC, t\dup\tE_2\wedge\tF \vdash M_1(x(ty))\dup\tD\\\B_2\vdash M_2\dup(\tE_1\vee\tE_2)\wedge\tF\end{array}}{\B_1, \B_2,x\dup\tG, y\dup\tC \vdash M_1(x(M_2y))\dup\tD}\]
By induction $\B_1\vdash \lambda z. M_1 z\dup\tB\to\tD$ and $t\dup\tE_1\wedge\tF\vdash \lambda z. t z\dup\tC\to\tA$ and  $t\dup\tE_2\wedge\tF\vdash \lambda z. t z\dup\tC\to\tA$.
Rule $(\vee L)$ derives $t\dup(\tE_1\vee\tE_2)\wedge\tF\vdash \lambda z. t z\dup\tC\to\tA$, so the application of rule $(C)$ to the third premise gives $\B_2 \vdash \lambda y. M_2y\dup\tC\to\tA$.

In the fourth case:
\[(\vee E)\quad\frac{\begin{array}{c}\B_1,\B_2,x\dup\tG, t\dup\tE_1\wedge\tF \vdash M_1(x(M_2t))\dup\tD~~~~\B_1,\B_2,x\dup\tG, t\dup\tE_2\wedge\tF \vdash M_1(x(M_2t))\dup\tD\\
y\dup\tC\vdash y\dup(\tE_1\vee\tE_2)\wedge\tF\end{array}}{\B_1, \B_2,x\dup\tG, y\dup\tC \vdash M_1(x(M_2y))\dup\tD}\]
 By induction on one of the first two premises $\B_1\vdash \lambda z. M_1 z\dup\tB\to\tD$ and $\B_2\vdash\lambda z. M_2 z\dup\tC\to\tA$.

 In the fifth case:
\[(\vee E)\quad\frac{\begin{array}{c}\B_1, t\dup\tE_1\wedge\tF \vdash M_1t\dup\tD~~~~\B_1, t\dup\tE_2\wedge\tF \vdash M_1t\dup\tD~~~~\B_2,x\dup\tG, y\dup\tC\vdash x(M_2y)\dup(\tE_1\vee\tE_2)\wedge\tF\end{array}}{\B_1, \B_2,x\dup\tG, y\dup\tC \vdash M_1(x(M_2y))\dup\tD}\]
The third premise with $x\dup\tA\to\tB\vdash x\dup\tG$ give $\B_2,x\dup\tA\to\tB, y\dup\tC\vdash x(M_2y)\dup(\tE_1\vee\tE_2)\wedge\tF$, so by Point~(\ref{first1}) \mbox{$\B_2\vdash\lambda z. M_2 z\dup\tC\to\tA$.} By Lemma \ref{simpleM2}(\ref{ss1}) $\B_2,x\dup\tA\to\tB, y\dup\tC\vdash x(M_2y)\dup(\tE_1\vee\tE_2)\wedge\tF$ implies $z\dup\tB\vdash z\dup(\tE_1\vee\tE_2)\wedge\tF$. The application of rule $(\vee E)$ to the first two premises and to $z\dup\tB\vdash z\dup(\tE_1\vee\tE_2)\wedge\tF$ derives $\B_1, z\dup\tB\vdash M_1 z\dup\tD$, which implies $\B_1\vdash \lambda z. M_1 z\dup\tB\to\tD$  by rule $(\to I)$.

In the sixth case:
\[(\vee E)\quad\frac{\begin{array}{c}\B_1,x\dup\tG, t\dup\tE_1\wedge\tF \vdash M_1(xt)\dup\tD~~~~\B_1,x\dup\tG, t\dup\tE_2\wedge\tF \vdash M_1(xt)\dup\tD\\\B_2, y\dup\tC\vdash M_2y\dup(\tE_1\vee\tE_2)\wedge\tF\end{array}}{\B_1, \B_2,x\dup\tG, y\dup\tC \vdash M_1(x(M_2y))\dup\tD}\]
The first and the second premise with $x\dup\tA\to\tB\vdash x\dup\tG$ give $\B_1,x\dup\tA\to\tB, t\dup\tE_1\wedge\tF \vdash M_1(xt)\dup\tD$ and \mbox{$\B_1,x\dup\tA\to\tB, t\dup\tE_2\wedge\tF \vdash M_1(xt)\dup\tD$.}
So Point (\ref{first4})  implies $\B_1\vdash \lambda z. M_1 z\dup\tB\to\tD$, $t\dup\tE_1\wedge\tF \vdash t\dup\tA$ and $t\dup\tE_2\wedge\tF \vdash t\dup\tA$.
The application of rule $(\vee E)$ to the last two statements and to the third premise derives $\B_2,y\dup\tC \vdash M_2y\dup\tA$,  so rule $(\to I)$ concludes the proof.

 (\ref{first3}). The Subject Expansion (Theorem \ref{setheorem}) gives  $\vdash\lambda xy. \id_1(x(\id_2y))\dup(\tM\to\ti)\to\tN\to\tk$. Corollary \ref{cor}(\ref{cor3}) implies\\ \centerline{$x\dup\tM\to\ti, y\dup\tN\vdash\id_1(x(\id_2y))\dup\tk$.}
Point (\ref{first2}) and Subject Reduction conclude the proof.
\end{proof}

\begin{definition}[Set of d-paths of an \fhi]\labelx{Ipd}
The set of d-paths of the \fhi\ \id\ (notation \Ip(\id)) is defined by:
\[\begin{array}{c}
\Ip(\lambda x.x)= \set\epsilon\quad \Ip(\lambda xy. \id_1(x(\id_2y)))= \set{\lp\q\mid\q\in \Ip(\id_2)}\cup\set{\rp\q\mid\q\in \Ip(\id_1)}.
\end{array}\]
\end{definition}

\begin{lemma}\labelx{nt1}
\begin{enumerate}
\item \labelx{nt11}
If $\vdash\id\dup\tA\to\tB$, where $\tA,\tB$ are both basic intersections  or basic unions in normal form, then
$\tA\md\tB$ and ${\Ip(\id)}\supseteq \Sp(\tA\md\tB)$, with $\Diamond=\wedge$ if $\tA,\tB$ are intersections and $\Diamond=\vee$ if $\tA,\tB$ are unions.
\item \labelx{nt13} If $\vdash \id \dup \tA \to\tA$ where $\tA$ is either a basic intersection or a basic union in normal form, then $\ag{\tA}{\Ip(\id)}$.
\end{enumerate}
\end{lemma}\spaziop
\begin{proof}
(\ref{nt11}).
By induction on $\id$.
If $\id=\lambda x.x$, then $x\dup\tA\vdash x\dup\tB$ by Corollary \ref{cor}(\ref{cor3}). This implies either $\tA=\tB$ or $\tA=\tM\wedge\tN$ and $\tB=\tM$ or $\tA=\ti$ and $\tB=\ti \vee\tk$ by Lemma \ref{var-un-in}. By definition either $\Sp(\tA\md\tB)=\set~$ or $\Sp(\tA\md\tB)=\set\epsilon$.
If $\id\not=\lambda x.x$, the following stronger statement is proved:\\[1mm]
{\em If $\vdash\id\dup\tA_l\to\tB_l$, where $\tA_l,\tB_l$ are normal types and either
intersections of arrows  or unions of arrows for all $l\in L$, then $\tA_l\md\tB_l$  and ${\Ip(\id)}\supseteq{\bigcup_{l\in L}\Sp(\tA_l\md\tB_l)}$ for all $l\in L$, with $\Diamond=\wedge$ if $\tA_l,\tB_l$ are intersections and $\Diamond=\vee$ if $\tA_l,\tB_l$ are unions.} \\[1mm]
If $\id\tob\lambda xy. \id_1(x(\id_2y))$, 
let $\tA_l=\bigwedge_{i\in I_l} (\tM^{(l)}_i\to\ti^{(l)}_i)$, $\tB_l=\bigwedge_{j\in J_l} (\tN^{(l)}_j\to\tk^{(l)}_j)$ (the proof for the
case the types are basic unions is similar). Theorem \ref{last} and Corollary \ref{cor}(\ref{cor3}) imply for all $l\in L$ and $j\in J_l$ there is $i_j\in I_l$ such that $x\dup \tM^{(l)}_{i_j}\to\ti^{(l)}_{i_j}, y\dup\tN^{(l)}_j\vdash \id_1(x(\id_2y))\dup\tk^{(l)}_j$.  Lemma \ref{first}(\ref{first3})  implies $\vdash \id_1\dup\ti^{(l)}_{i_j}\to\tk^{(l)}_j$ and $\vdash \id_2\dup\tN^{(l)}_j\to\tM^{(l)}_{i_j}$ for all $l\in L$ and $j\in J_l$. By induction for all $l\in L$ and $j\in J_l$:\\
 \centerline{$\ti^{(l)}_{i_j}\muu\tk^{(l)}_j$ and ${\Ip(\id_1)}\supseteq{\bigcup_{l\in L}\bigcup_{j\in J_l}\Sp(\ti^{(l)}_{i_j}\muu\tk^{(l)}_j)}$
}
 \centerline{$\tN^{(l)}_j\mi\tM^{(l)}_{i_j}$  and ${\Ip(\id_2)}\supseteq{\bigcup_{l\in L}\bigcup_{j\in J_l}\Sp(\tN^{(l)}_j\mi\tM^{(l)}_{i_j})}$
 }
By definition of $e$ (Figure \ref{Spd}) $\bigcup_{l\in L}\Sp(\tA_l\mi\tB_l)=\bigcup_{l\in L}\bigcup_{j\in J_l}(\lp\cdot\Sp(\tN^{(l)}_j\mi\tM^{(l)}_{i_j})\;\cup \rp \cdot\Sp(\ti^{(l)}_{i_j}\muu\tk^{(l)}_j))
\subseteq\Ip(\id)$.
\\
(\ref{nt13}). By induction on $\id$. If $\id = \lambda x.x$, then $\Ip(\id)= \set{\epsilon}$ and the proof is immediate. \\
If $\id\tob\lambda xy. \id_1(x(\id_2y))$ let $\tA = \bigwedge_{i \in I} (\tM_i \to \ti_i)$ (the proof for the case of the union is similar). Let $I=J\cup H$ , with $J\cap H = \emptyset$, and $J$ be the maximum subset of $I$ such that
$\der{}{\id}{ (\tM_j \to \ti_j)\to\tM_j \to \ti_j}$ for all $j\in J$.\\
The proof starts by showing that $J$ can not be empty. Theorem \ref{last} assures that for all $i\in I$ there is $j_i\in I$ such that $\der{}{\id}{(\tM_{j_i} \to \ti_{j_i})\to\tM_i \to \ti_i}$. If there are $i_1,\ldots,i_n$ such that $\der{}{\id}{(\tM_{i_l} \to \ti_{i_l})\to\tM_{i_{l+1}} \to \ti_{i_{l+1}}}$ for $1\leq l\leq n-1$ and $\der{}{\id}{(\tM_{i_n} \to \ti_{i_n})\to\tM_{i_1} \to \ti_{i_1}}$, then also $\der{}{\id}{(\tM_{i_l} \to \ti_{i_l})\to\tM_{i_{l}} \to \ti_{i_{l}}}$ for $1\leq l\leq n$, since $\id\tob\id\circ\id$.\\
 From $\der{}{\id}{(\tM_j \to \ti_j)\to\tM_j \to \ti_j}$ Corollary \ref{cor}(\ref{cor3})
and  Lemma \ref{first}(\ref{first2}) give $\id_2 \dup \tM_j \to\tM_j$ and $\id_1 \dup \ti_j \to\ti_j$. By induction $\ag{\tM_j}{\Ip(\id_2)}$ and $\ag{\ti_j}{\Ip(\id_1)}$ for all $j \in J$, which imply
  $\ag{\bigwedge_{j\in J} (\tM_j \to \ti_j)}{\Ip(\id)}$. Moreover by assumption for all $h\in H$ there is
$j_h \in J$ such that $\id\dup (\tM_{j_h} \to \ti_{j_h} ) \to \tM_{h} \to \ti_{h}$. Point (\ref{nt11}) implies $\tM_{j_h} \to \ti_{j_h} \md \tM_{h} \to \ti_{h}$ and
$\Ip(\id)\supseteq\Sp(\tM_{j_h} \to \ti_{j_h} \md \tM_{h} \to \ti_{h}) $.  The second erasure rule gives $\tA\red\bigwedge_{j\in J} (\tM_j \to \ti_j)$. Therefore $\tA$ would not be a normal type. So $H$ must be empty.
\end{proof}

\begin{theorem}\labelx{ntT1}
\begin{enumerate}
\item \labelx{ntT11} If  $\vdash\id\dup\bigwedge_{j\in J} \ti_j\to\bigwedge_{i\in I} \ti_i$ and $J \subset I$, then $\bigwedge_{i\in I} \ti_i$ is not a normal type.
\item \labelx{ntT12} If $\vdash\id\dup\bigvee_{i\in I} \tO_i\to\bigvee_{j\in J} \tO_j$ and $J \subset I$, then $\bigvee_{i\in I} \tO_i$ is not a normal type.
\end{enumerate}
\end{theorem}\spaziop
\begin{proof}
(\ref{ntT11}). Assume ad absurdum that $\bigwedge_{i\in I} \ti_i$ is a normal type. Let $I=K\cup H$, with $K\cap H=\emptyset$, and $H$ be the maximum subset of $I$ such that for all $h\in H$ there is
$k_h \in K$ such that $\id\dup (\tM_{k_h} \to \ti_{k_h} ) \to \tM_{h} \to \ti_{h}$.
 Notice that by construction $H\supseteq I-J$, therefore $H$ can not be empty. By Lemma  \ref{nt1}(\ref{nt11})  $\tM_{k_h} \to \ti_{k_h} \md \tM_{h} \to \ti_{h}$ and
$\Ip(\id)\supseteq\Sp(\tM_{k_h} \to \ti_{k_h} \md \tM_{h} \to \ti_{h}) $.  Moreover by assumption $\der{}{\id\,}{\, \ti_k\to \ti_k}$ for all $k\in K$. Lemma  \ref{nt1}(\ref{nt13}) implies  $\ag{\bigwedge_{k\in K}\ti_k}{\Ip(\id)}$. The first erasure rule gives $\bigwedge_{i\in I} \ti_i\red\bigwedge_{k\in K} \ti_k$, proving that $\bigwedge_{i\in I} \ti_i$ is not a normal type.


(\ref{ntT12}). Similar to the proof of (\ref{ntT11}), using the last erasure rule.
\end{proof}

\begin{theorem} \labelx{discon}
Let $\bigwedge_{i\in I}(\bigvee_{h\in H_i}\tO^{(i)}_h)\approx\bigwedge_{j\in J}(\bigvee_{k\in K_j}\tP^{(j)}_k)$ and both types be normal. Then $I=J$, $H_i=K_i$ and $\tO^{(i)}_h\approx\tP^{(i)}_h$ for all $h\in H_i$ and $i\in I$.
\end{theorem}\spaziop
\begin{proof}
Let $<P,P^{-1}>$ prove the isomorphism and let $\ti_i=\bigvee_{h\in H_i}\tO^{(i)}_h$ and $\tk_j=\bigvee_{k\in K_j}\tP^{(j)}_k$.\\ Assume ad absurdum that $I \subset J$.
By Theorem \ref{last} for all $j\in J$  there is $i_j\in I$ such that
$\vdash P\dup\ti_{i_j}\to\tk_j$ and for $i_{j}\in I$ there is  $j_{i_j}\in J$ such that $\vdash P^{-1}\dup\tk_{j_{i_j}}\to\ti_{i_j}$. This implies $\vdash P\circ P^{-1}\dup\tk_{j_{i_j}}\to \tk_j$ and, for cardinality reasons, there are $i ,\;j$ such that $j_{i_j}\not=j$. This, together with $\vdash P\circ P^{-1}\dup\bigwedge_{j\in J}\tk_{j}\to \bigwedge_{j\in J}\tk_j$, gives $\vdash P\circ P^{-1}\dup\bigwedge_{j\in J'}\tk_{j}\to \bigwedge_{j\in J}\tk_j$ for some $J'\subseteq I$. By Theorem \ref{ntT1}(\ref{ntT11}) $\bigwedge_{j\in J}\tk_j$ is not a normal type. Then $I=J$ and $j_{i_j}=j$. Therefore the indexes can be chosen to get $\ti_i\approx\tk_i$ for all $i\in I$.\\
Assume ad absurdum that $K_i \subset H_i$.
By Theorem \ref{dis} for all $h\in H_{i}$ there is  $k_h\in K_{i}$ such that
$\vdash P\dup\tO^{(i)}_{h}\to\tP^{(i)}_{k_h}$ and for all $k_h\in K_{i}$ there is $h_{k_h}\in H_{i}$ such that $\vdash P^{-1}\dup\tP^{(i)}_{k_h}\to\tO^{(i)}_{h_{k_h}}$. This implies
$\vdash P^{-1}\circ P\dup\tO^{(i)}_{h}\to \tO^{(i)}_{h_{k_h}}$ and there are $h,\;k$ such that $h_{k_h}\not=h$. This fact, together with $\vdash P^{-1}\circ P\dup\bigvee_{h\in H_i}\tO^{(i)}_{h}\to \bigvee_{h\in H_i}\tO^{(i)}_{h}$, gives $\vdash P^{-1}\circ P\dup\bigvee_{h\in H_i}\tO^{(i)}_{h}\to \bigvee_{h\in H'}\tO^{(i)}_{h}$ for some $H'\subseteq K_i$. By Theorem \ref{ntT1}(\ref{ntT12})  $\bigvee_{h\in H_i}\tO^{(i)}_{h}$ is not a normal type.
Then $H_i=K_i$ and $h_{k_h}=h$. Therefore the indexes can be chosen to get $\tO^{(i)}_h\approx\tP^{(i)}_h$ for all $h\in H_i$ and $i\in I$.
\end{proof}

\section{Conclusion}\spaziot
This paper introduces a system with intersection and union types for linear $\lambda$-terms. The system enjoys subject conversion owing to the linearity restriction. The types that can be derived for
the $\lambda$-terms proving type isomorphism are studied. 
A main achievement of this paper is the definition of rules to reduce types to normal form, while preserving isomorphism. These rules are the building blocks for characterising type isomorphism by means of a syntactic equivalence relation between types. This characterisation is the content of \cite{CDMZ13}, where all proofs given in the present paper are omitted. The present paper and \cite{CDMZ13} can be considered as the first and the second part of a unique work. \vspace{-3.5mm}

\paragraph{Acknowledgments}
The authors gratefully thank the anonymous referees  for their useful remarks.

\vspace{-4mm}

\bibliographystyle{eptcs}
\bibliography{biblio}

\end{document}